\documentclass[11pt]{article}
\usepackage{amsmath,amssymb}
\usepackage{amsthm}
\usepackage{graphicx}
\usepackage{typearea}
\typearea{13}

\numberwithin{equation}{section}

\newcommand{\e}{\epsilon} 

\newcommand{\la}{\lambda} 

\newcommand{\La}{\Lambda}

\newcommand{\Om}{\Omega}

\newcommand{\Ncal}{\mathcal{N}}

\newcommand{\Tcal}{\mathcal{T}}

\newcommand{\Rbb}{\mathbb{R}}

\newcommand{\lp}{\left(}
\newcommand{\rp}{\right)}
\newcommand{\lc}{\left\{}
\newcommand{\rc}{\right\}}

\begin{document}


\renewcommand{\thefootnote}{\fnsymbol{footnote}}
\setcounter{page}{0}
\thispagestyle{empty}
\begin{flushright} OU-HET 904 \end{flushright} 

\vskip3cm
\begin{center}
{\LARGE {\bf Surface Operators from M-strings}}  
\vskip1.5cm
{\large 
{\bf Hironori Mori\footnote{\href{mailto:hiromori@het.phys.sci.osaka-u.ac.jp}{hiromori@het.phys.sci.osaka-u.ac.jp}}
and
\bf Yuji Sugimoto\footnote{\href{mailto:sugimoto@het.phys.sci.osaka-u.ac.jp}{sugimoto@het.phys.sci.osaka-u.ac.jp}}} 

\vskip1cm
\it Department of Physics, Graduate School of Science, Osaka University, \\ Toyonaka, Osaka 560-0043, Japan} 
\end{center}

\vskip1cm
\begin{abstract} 
It has been found that surface operators have a significant role in AGT relation. This duality is an outstanding consequence of M-theory, but it is actually encoded into the brane web for which the topological string can work. From this viewpoint, the surface defect in AGT relation is geometrically engineered as a toric brane realization. Also, there is a class of the brane configuration in M-theory called M-strings which can be translated into the language of the topological string. In this work, we propose a new M-string configuration which can realize AGT relation in the presence of the surface defect by utilizing the geometric transition in the refined topological string.
\end{abstract}

\renewcommand{\thefootnote}{\arabic{footnote}}
\setcounter{footnote}{0}

\vfill\eject

\tableofcontents

\section{Introduction and summary}
A surface operator which we would like to study is a non-local defect with codimension-2 in four-dimensional gauge theories. This is a kind of disorder operators, like a 't Hooft line, which cannot be expressed using fundamental fields in the theory rather should be specified by singular behaviors of them in approaching to the operators. This type of the operator in 4d is really special because the codimension is identical to the dimension supported by the operator, which is similar to the well-known fact that a gauge field has a classical (anti)self-dual solution in the 4d gauge theory. The full characteristics of the surface operator as the disorder one is less understood than those of the order operator like a Wilson line, though lots of attempts to completely classify the surface operator have been archived (e.g., see \cite{Gukov:2006jk, Gukov:2008sn, Gomis:2007fi, Drukker:2008wr, Koh:2008kt, Koh:2009cj, Gaiotto:2009fs, Gaiotto:2012xa, Gomis:2014eya, Chen:2014rca, Maruyoshi:2016caf, Ito:2016fpl} and a comprehensive review \cite{Gukov:2014gja}).
Nevertheless, the surface operator can be used to investigate non-perturbative properties of the theories and uncover nontrivial consequences in the framework of the duality. In AGT correspondence \cite{Alday:2009aq}, the insertion of the surface operator on the gauge theory side is interpreted as introducing a degenerate operator on the Riemann surface \cite{Alday:2009fs}, and this fact actually provides considerable aspects which we might not reach unless the surface operator is taken into account.


In this paper, we keep our attention on AGT relation in the presence of the surface operator. This relation is originally encoded into M-theory where the 6d $\Ncal = ( 2, 0 )$ superconformal field theory as the worldvolume theory on multiple M5-branes is compactified on a four-sphere and a Riemann surface. The surface operator which we mainly consider is naturally viewed as the boundary of a M2-brane ending on those M5's which appear as a degenerate operator on the Riemann surface. Further, it is known that with utilizing string duality, AGT correspondence is embedded into the system of $( p, q )$-fivebranes where we can apply the topological string computation. On the gauge theory side, the topological string partition function just produces the 5d uplift of a Nekrasov partition function which can really be reduced to the 4d one on the sphere. On the CFT side, the structure of the web digram for the $( p, q )$-fivebranes is mapped to the geometry of the Riemann surface. In addition, the surface operator is encoded in this picture from the insertion of an extra D3-brane into the web diagram which corresponds to a Lagrangian brane in the topological string. This relation through the web diagram in string theory is currently thought of as the 5d version of AGT correspondence \cite{Awata:2009ur}.

\begin{figure}[t] 
	\begin{center}
	\includegraphics[width=10cm,bb= 160 200 680 440,clip]{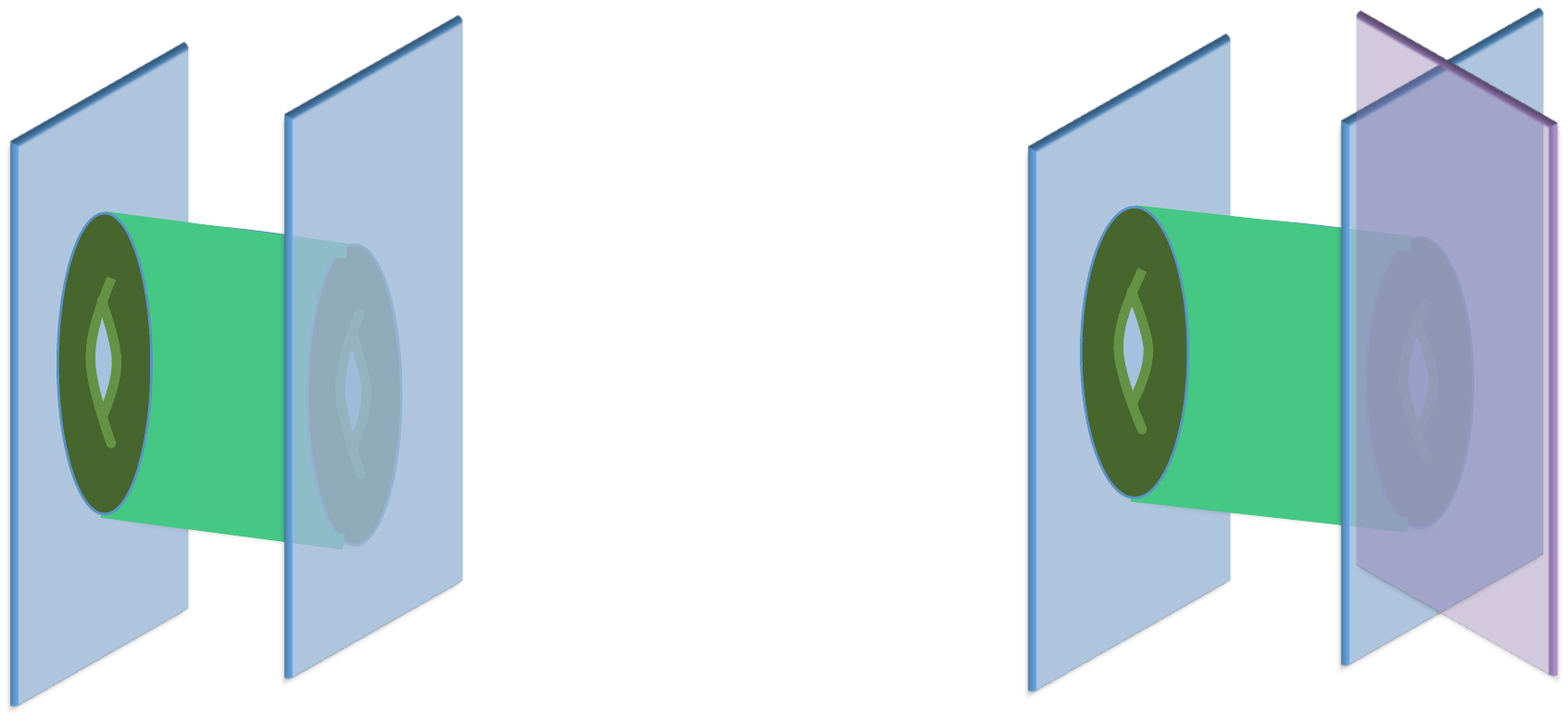}
	\caption{The left side is the original M-strings where multiple M2's (green) are suspended between two parallel M5's (blue). The M-string is expressed by a torus on the M5. The right side is our new proposal for the M-strings with an additional M5 (purple).}
	\label{proposal}
	\end{center}
\end{figure}

There is another M-theory setup to understand the contribution from the M2 in more efficient way. This is named a M-string \cite{Haghighat:2013gba} which is the boundary of the M2 suspended between two parallel M5's (the left side of Figure \ref{proposal}). It can be immediately seen that the M-string configuration is rewritten as the $( p, q )$-fivebrane web diagram again by using the power of string duality, and, accordingly, the refined topological string is used to calculate the partition function of the M-strings. It does not seem so simple that AGT correspondence is interpreted as the language of the M-strings, however, in this work, we propose the M-string configuration which appropriately engineers AGT correspondence in the presence of the surface operator. This new system includes an extra M5 in addition to the original M-strings, i.e., the M2-M5 system on the flat background (the right side of Figure \ref{proposal}), and this new M5 can geometrically engineer the surface operator via the geometric transition in the topological string \cite{Gopakumar:1998ii, Gopakumar:1998ki, Gopakumar:1998jq}. In other words, in the M-strings, the surface operator is realized not from the M2 in the usual M-theory construction of AGT relation, but from the M5 intersecting with one of the original M5 in M-strings. We will discuss this point based on T-duality in Section \ref{Mstrings}.

The rest of this paper is organized as follows. In Section \ref{Surface}, we shortly review how the surface operator is engineered via the geometric transition of the bubbling Calabi-Yau in the refined topological string. In Section \ref{Mstrings}, it is explained that our M-string configuration with an additional M5-brane can produce the surface operator in AGT correspondence. We exactly compute the contribution of this new M5 based on the refined topological vertex, which turns to be equivalent to the elliptic genus of the 2d $\Ncal=(2,2)$ $\text{U}(k)$ theory as expected from the M-theory point of view. Finally, we summarize the results and comment on several outlooks in Section \ref{Sum}. The definitions of the refined topological vertex and the detailed computation of the M-strings are included in Appendix \ref{Convension}.

\section{Surface operators in topological string}\label{Surface}

\subsection{Degenerate limit in AGT correspondence}
In AGT correspondence \cite{Alday:2009aq}, the presence of the surface operators in the 4d gauge theories is expressed just as the insertion of the degenerate operators on the Riemann surface into the correlation function in CFTs \cite{Alday:2009fs}. The surface operator in AGT relation is usually originated from inserting the M2-brane on M5-branes by hand. Instead, we can built the situation in the presence of the surface operator by taking the specific limit via AGT relation. On the CFT side, a regular puncture in this limit turns to be a degenerate operator on the Riemann surface, which means that the corresponding surface operator just arises in the 4d gauge theory. It is physically unclear in the field theory viewpoint why this may occur only with tuning the parameters to such special values. However, we can give somehow intuitive explanation for this construction of the surface operators \cite{Dimofte:2010tz, Taki:2010bj} by using the geometric engineering in the topological string theory \cite{Gopakumar:1998ii, Gopakumar:1998ki, Gopakumar:1998jq}. Let us briefly review this prescription.

\begin{figure}[t] 
	\begin{center}
	\includegraphics[width=9cm,bb= 90 40 720 530,clip]{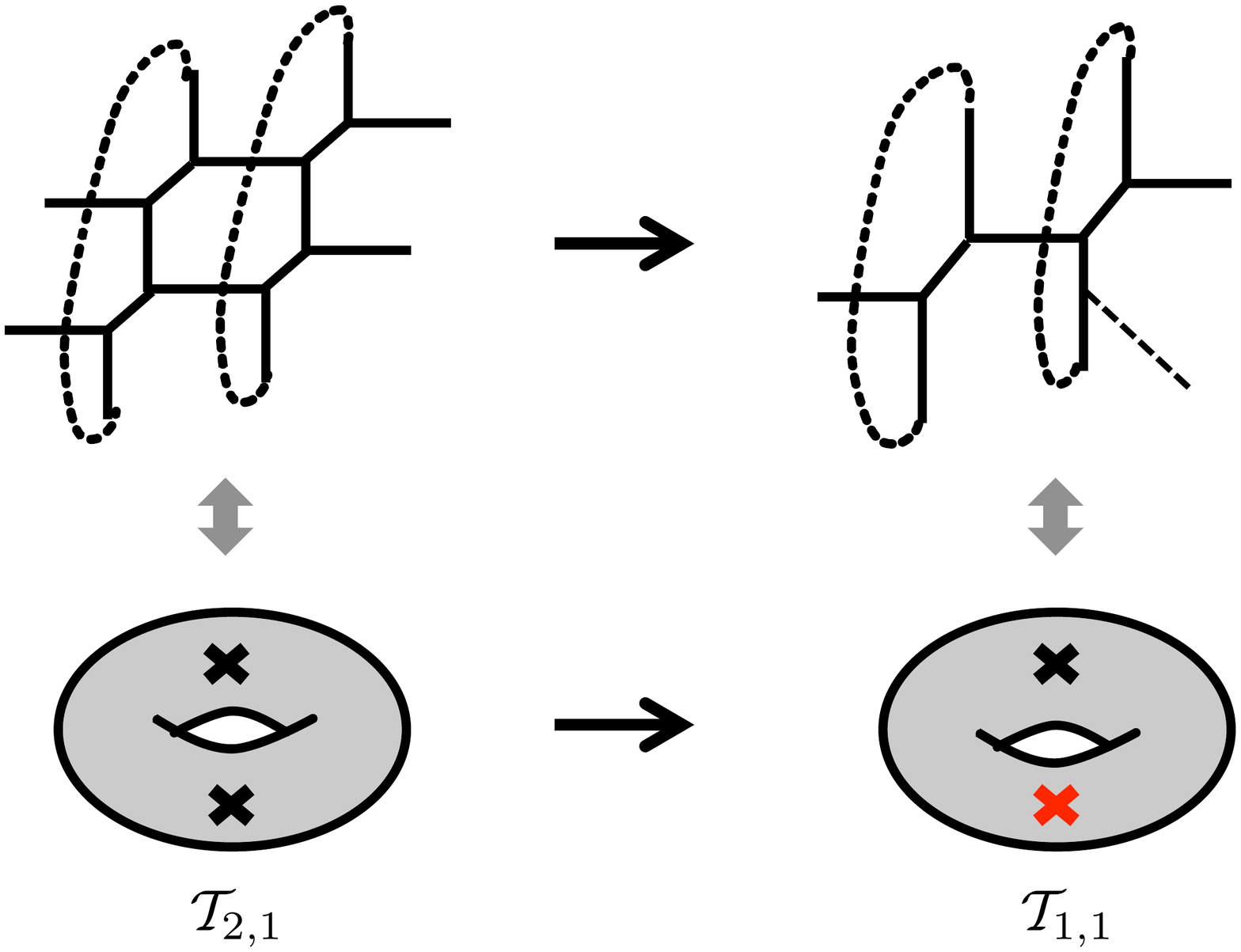}
	\caption{$\Tcal_{2, 1}$ and its degeneration limit to $\Tcal_{1, 1}$ with a degenerate operator (a red cross). The web diagrams on the left and the right of the upper line correspond to the $\hat{\text{A}}_1$ quiver gauge theory and the $\Ncal=2^\ast$ SU$(2)$ gauge theory with a surface operator, respectively.}
	\label{deglim}
	\end{center}
\end{figure}

We focus on one example from now on which will be used in the next section. The starting point is the so-called $\hat{\text{A}}_1$ quiver gauge theory, a certain 4d $\text{SU}(2)$$\times$$\text{SU}(2)$ gauge theory, which has the description of the $(p,q)$-fivebrane web diagram. The corresponding CFT is on the torus with two primary fields (punctures) denoted by $\Tcal_{2, 1}$ (the left side of Figure \ref{deglim}), where $\Tcal_{n, g}$ represents the theory on the Riemann surface with $n$ punctures and $g$ handles. Then, the degeneration limit of $\Tcal_{2, 1}$ which we consider to produce the surface operator insertion is,
\begin{align}
a = - m_1 = m_2 + \epsilon_2,
\label{dlim}
\end{align}
where, in the gauge theory language, $a$ is a SU$(2)$ gauge holonomy and $m_{1, 2}$ are masses of the matters corresponding to scaling dimensions in the CFT. $\epsilon_2$ is one of $\Om$-deformation parameters. This is because, on flat background $\Rbb^4$, the surface operator along $\Rbb^2 \subset \Rbb^4$ is affected by an U$(1)$ rotation labelled by $\e_2$. Then, $\Tcal_{2, 1}$ in the limit \eqref{dlim} results in $\Tcal_{1, 1}$ with a degenerate operator, that is, the theory on the torus with one puncture and one degenerate field (the right side of Figure \ref{deglim}). The dictionary of AGT relation tells us that $\Tcal_{1, 1}$ is mapped to the $\Ncal = 2^\ast$ $\text{SU}(2)$ gauge theory with a surface operator on the gauge theory side. This 4d theory can actually be built also from the web diagram with the insertion of a Lagrangian brane in the sense of the topological string. The reason for the appearance of the surface operator simply from the degenerate limit now becomes obvious in terms of the geometric transition in the web diagram, which we will explain in the next subsection.

\subsection{Geometric engineering}
\begin{figure}[t] 
	\begin{center}
	\includegraphics[width=11cm,bb= 70 170 680 410,clip]{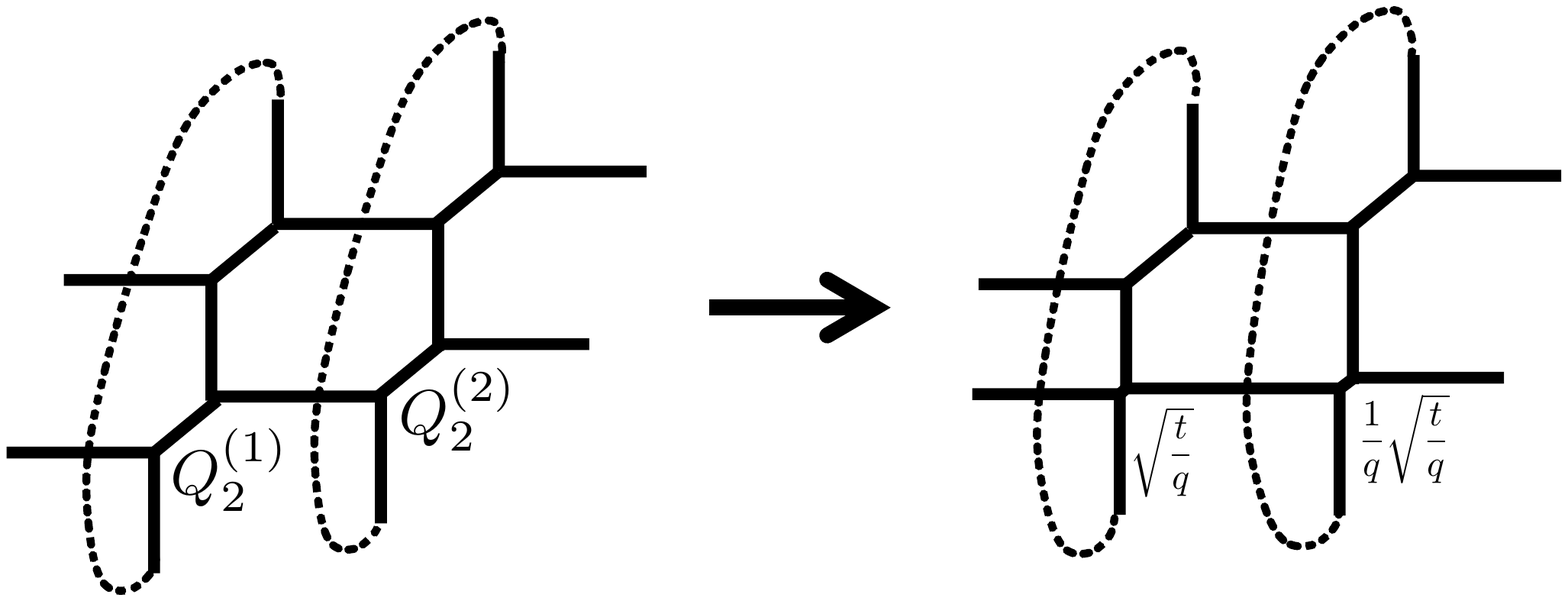}
	\caption{The degenerate limit of the K\"ahler parameters.}
	\label{degenerate}
	\end{center}
\end{figure}

\begin{figure}[t] 
	\begin{center}
	\includegraphics[width=11cm,bb= 90 190 690 420,clip]{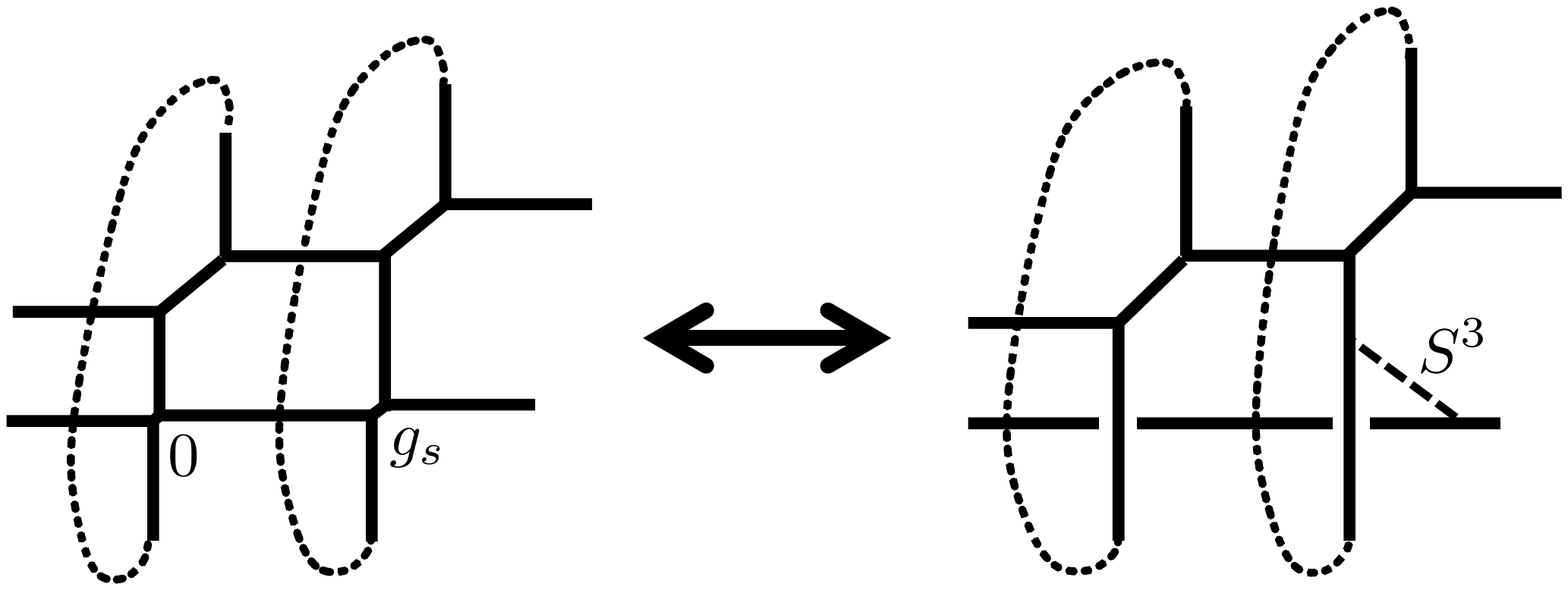}
	\caption{The geometric transition for the mergence of a toric brane corresponding to a surface operator.}
	\label{geosurface}
	\end{center}
\end{figure}

The geometric transition (or the open/closed duality) provides the connection between a brane and a background geometry. Let us begin the argument with the web diagram shown in the left side of Figure \ref{deglim} known as the bubbling Calabi-Yau \cite{Gomis:2006mv, Gomis:2007kz}. On this diagram, K\"ahler parameters $Q_2^{( 1 )}$ and $Q_2^{( 2 )}$ which essentially characterize sizes of the branes are set in the standard way (the left side of Figure \ref{degenerate}) and reduce in the limit \eqref{dlim} to
\begin{align}
    \begin{aligned}
    Q_2^{(1)} &\to \sqrt{\frac{t}{q}},\\ 
    Q_2^{(2)} &\to \frac{1}{q} \sqrt{\frac{t}{q}}, 
    \end{aligned}
    \label{limK}
\end{align}
where $t := \exp ( 2 \pi i \epsilon_1 )$ and $q := \exp ( - 2 \pi i \epsilon_2 )$. We now consider the meaning of this specialization for the unrefined case ($\epsilon_1 + \epsilon_2 = 0$). That $Q_2^{(1)}$ becomes 1 implies that the brane associated with this K\"ahler parameter reduces to the zero size, in contrasts, the nontrivial $Q_2^{(2)}$ value turns to be the topological string coupling $g_s$ (the left side of Figure \ref{geosurface}). Then, we apply the geometric transition with this specific limit which translates the branes into the geometry. The diagonal one attached on the right leg in the left side of Figure \ref{geosurface} represents a $\mathbb{CP}^1$ with size $g_s$, and after the transition this goes to the insertion of a Lagrangian 3-cycle $S^3$ depicted by the dotted line in the right side of Figure \ref{geosurface}. The other attached on the left leg in the left side of Figure \ref{geosurface} shows the zero size $\mathbb{CP}^1$, which leads to the Lagrangian 3-cycle without the brane. According to the geometric transition, the toric brane realization of the surface operator in the closed topological string with the degenerate limit can be nicely explained by the open string topological string expressed as the insertion of the dotted line in Figure \ref{geosurface}. The elliptic uplift of the unrefined version of the degenerate limit is actually \eqref{limK} which can be verified from the CFT via AGT relation.

\begin{figure}[t] 
	\begin{center}
	\includegraphics[width=4.5cm,bb= 290 150 550 410,clip]{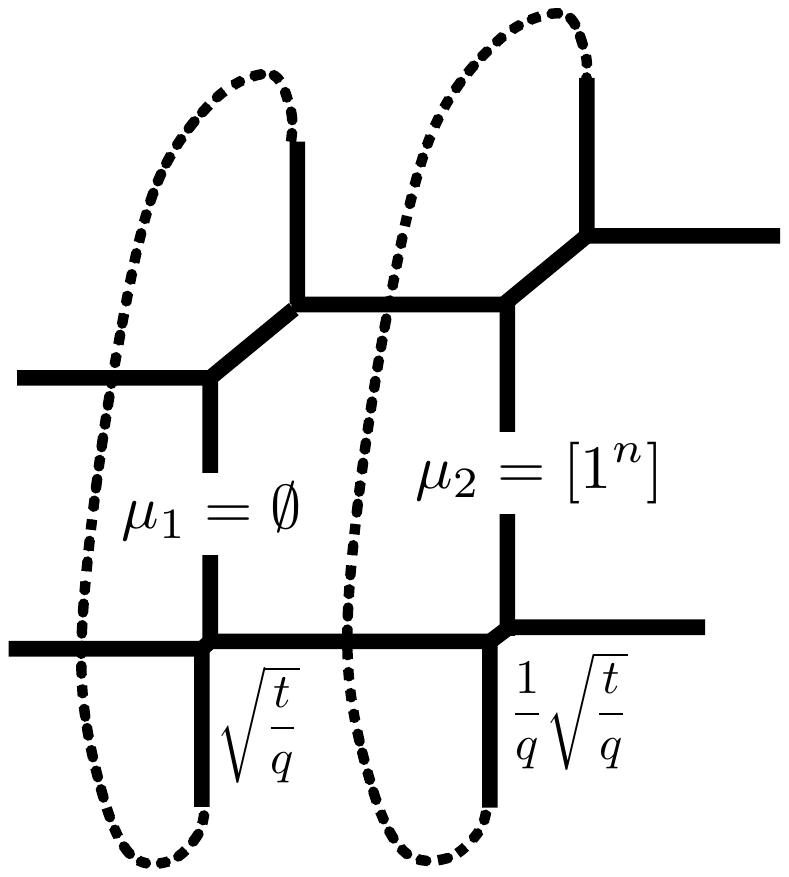}
	\caption{The restriction on the web diagram to get the nontrivial partition function.}
	\label{gluelim}
	\end{center}
\end{figure}

To justify this construction of the surface operator quantitatively, we move to the partition function of this web diagram which can be computed based on the refined topological vertex \cite{Iqbal:2007ii}. The consequence of the specialization \eqref{limK} affects it as follows. After doing the limit \eqref{limK}, this partition function contains the factor
\begin{align}
\prod_{( i, j ) \in \mu_1} ( 1 - t^{i - 1} q^{- j + 1} )
\prod_{( i, j ) \in \mu_2} ( 1 - t^{i - 1} q^{- j + 2} ),
\end{align}
where $\mu_{1, 2}$ are Young tableaux assigned in gluing the building blocks of the web diagram. It can be easily seen that this vanishes if $\mu_1$ includes even at most one box and $\mu_2$ has the box on the site $( i, j ) = ( 1, 2 )$. Therefore, we just obtain the nontrivial result only when we restrict the case (Figure \ref{gluelim})
\begin{align}
    \begin{aligned}
    \mu_1 &= \emptyset, \\ 
    \mu_2 &= [ 1^n ] \hspace{1em} ( n = 0, 1, 2, \cdots ), 
    \end{aligned}
\end{align}
where $[ 1^n ]$ is the Young tableau of a column with $n$ boxes. With this limitation, the refined topological string partition function in the 4d reduction turns to be the Nekrasov partition function in the presence of the surface operator, called the ramified instanton partition function, for the $\Ncal=2^\ast$ $\text{SU}(2)$ gauge theory. Also, it can be checked that it just matches the conformal block for the degenerate operator in $\Tcal_{1, 1}$ \cite{Taki:2010bj}. Note that this prescription can be straightforwardly applied to more general bubbling Calabi-Yau's, but we still keep attention to the simplest case shown above for avoiding complexity of calculations.

\section{M-strings calculation}\label{Mstrings}
In the previous section, we reviewed the topological string realization of the gauge theory with the surface operator. Then, it has been found that we can calculate the ramified partition function by using the refined open topological string on the certain non-compact Calabi-Yau manifold \cite{Aganagic:2003db, Iqbal:2007ii}. 
It has been suggested that M-strings \cite{Haghighat:2013gba, Haghighat:2013tka} has the dual picture of string theory which can be encoded into a $( p, q )$-web diagram, and its partition function is obtained also by the topological string theory.
\par
In this section, we consider the M-strings realization for this refined open topological string theory. Then, we will find that, by inserting one more M5 brane to the originai M-strings set up, this topological string theory is dual to the M-strings set up. Then, we calculate the partition function of the M-strings by using the refined topological vertex formalism.

\subsection{M-strings}
\begin{figure}[t]
\centering
    \includegraphics[width=8cm]{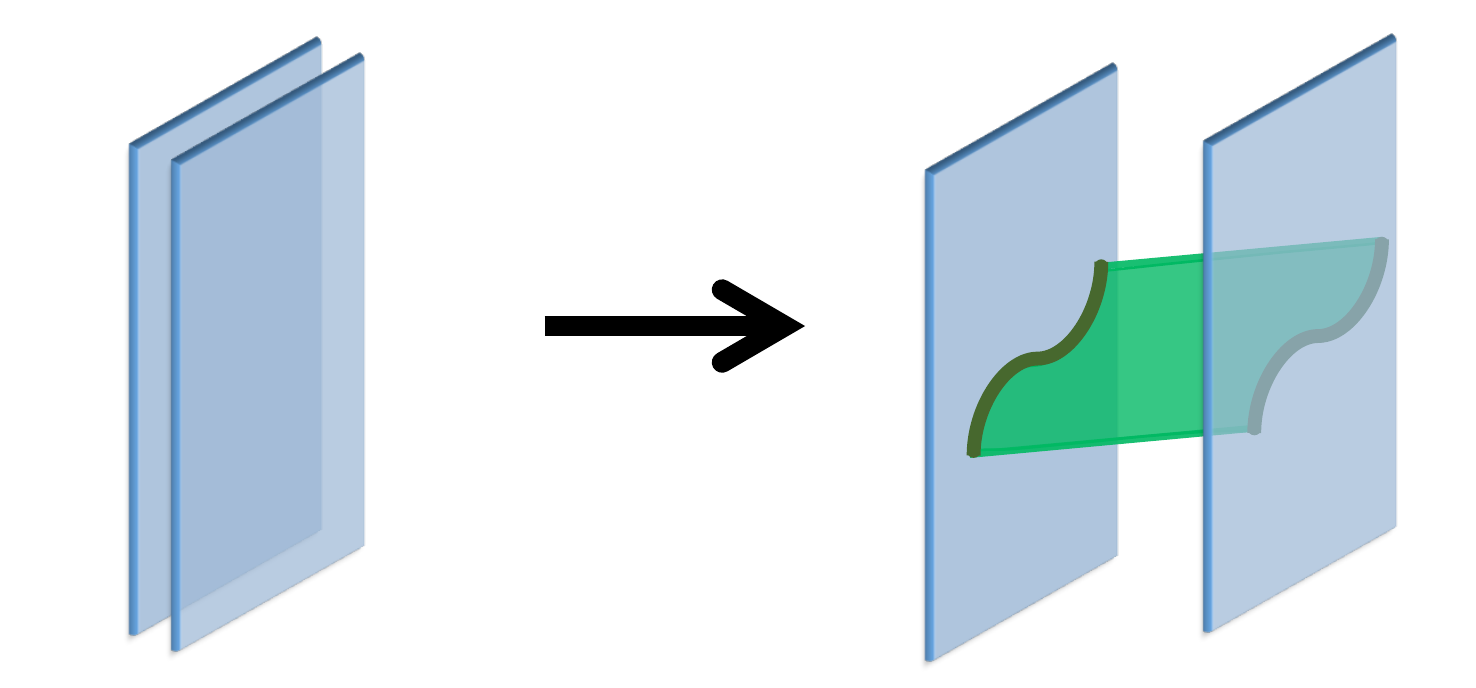}\vspace{1.5em}
    \begin{tabular}{|c||c|c|c|c|c|c|c|c|c|c|c|} \hline
     M-theory & $X_{0}$ & $X_{1}$ & $X_{2}$ & $X_{3}$ & $X_{4}$ & $X_{5}$ & $X_{6}$ & $X_{7}$ 
     & $X_{8}$ & $X_{9}$ & $X_{10}$  \\ \hline \hline
    $2$ M5 & 
	$\circ$ & $\circ$ & $\circ$ & $\circ$ & $\circ$ & $\circ$  &  &  &  & & \\ \hline
    $k$ M2 & 
	$\circ$ & $\circ$ &  &  &  &  & $\circ$ &  &  & & \\ \hline
    $\text{A}_{N-1}$ ALE &  &  &  &  &  &  &  & $\circ$ & $\circ$ & $\circ$& $\circ$ \\ \hline
    \end{tabular}
    \vspace{1em}
    \caption{The M-strings. The blue sheets and the green sheets are the M5-branes and the M2-branes, respectively. Their boundaries which are denoted by the dark green lines are the M-strings. We set the background geometry to be the $\text{A}_{N-1}$ ALE space.}
    \label{M-strings}
\end{figure}

To begin with, we review the M-strings.
Let us consider $M$ M5-branes in the $\text{A}_{N-1}$ ALE space and separate them by introducing $k$ M2-branes (Figure \ref{M-strings}). Then, we can see the two-dimensional object in their boundaries, so-called ``M-strings."
Since there are two kinds of branes on orbifold, supersymmetry which the M-strings have is $\mathcal{N}=(0,4)$.  Then, we consider the mass deformations in order to connect this M-strings set up to the topological string.
\par
Now we consider the duality chain. We compactify the $X_{0}$ and $X_{1}$ directions. In order to connect this brane set up to topological string theory, we fiber $\mathbb{R}^4$ over $X_{0}$, where $\mathbb{R}^4$ is defined as $(X_{2},X_{3},X_{4},X_{5})$ directions\footnote{In order to preserve supersymmetry, we also have to fiber $\text{A}_{N-1}$ geometry $(w_{1},w_{2}) \in \mathbb{C}^2$ around this circle. More discussions can be seen in \cite{Haghighat:2013gba, Haghighat:2013tka}.}, and we identify $\mathbb{R}^4$ with coordinates $(z_{1},z_{2}) \in \mathbb{C}^2$.
Then, as we go around the circle in the $X_{0}$ direction, we twist $\mathbb{R}^4$ by the action of $\text{U}(1)_{\epsilon_{1}}\times \text{U}(1)_{\epsilon_{2}}$,
\begin{eqnarray}
\text{U}(1)_{\epsilon_{1}}\times \text{U}(1)_{\epsilon_{2}}
:(z_{1},z_{2}) 
\mapsto
(e^{2\pi\mathrm{i}\epsilon_{1}}z_{1},e^{2\pi\mathrm{i}\epsilon_{2}}z_{2}).
\end{eqnarray}
\par
\begin{figure}[t]
\centering
    \includegraphics[width=13cm]{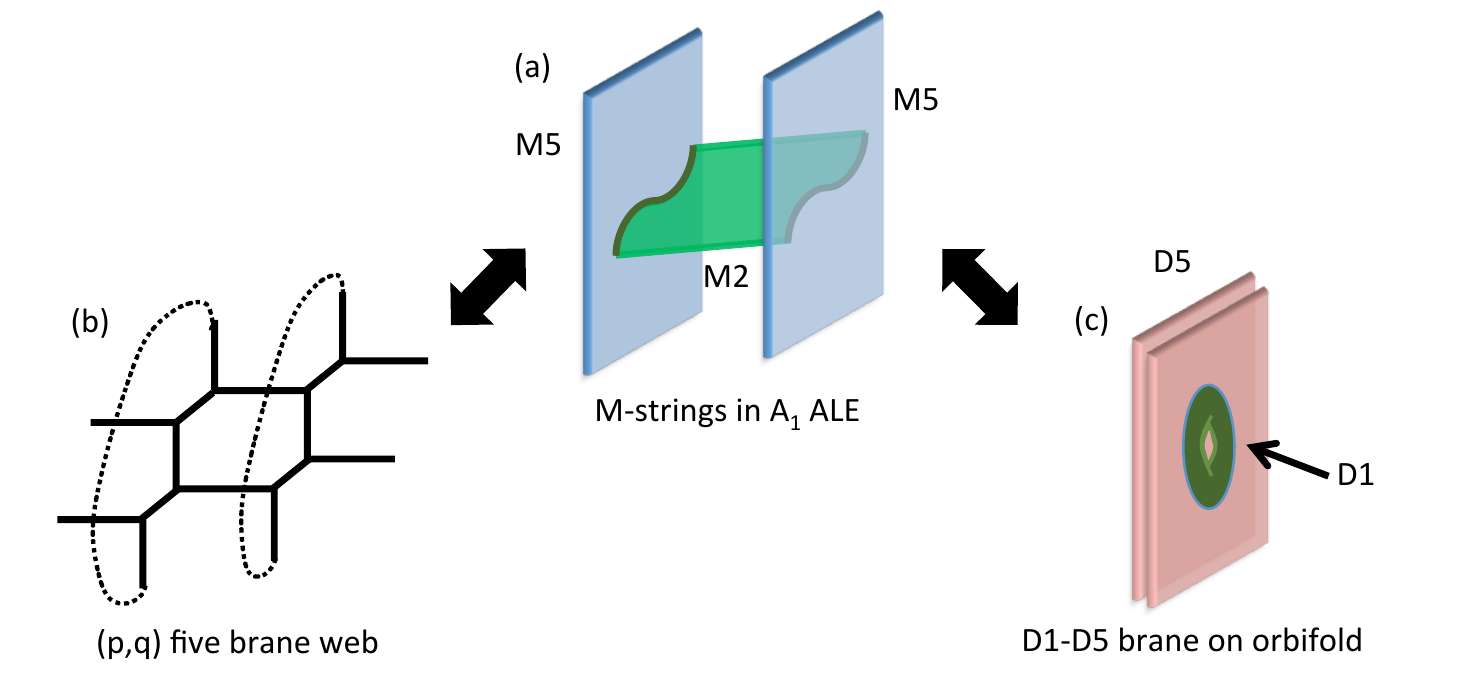}
    \caption{The duality chain for $M=2$ and $N=2$. The red sheets are the D5 branes which correspond to the $\text{A}_{N-1}$ ALE space.}
    \label{equivalence}
\end{figure}

Then, we take the $X_{1}$ direction as the M-theory circle and perform T-duality along the $X_{7}$ direction which is the part of the $\text{A}_{N-1}$ ALE space. As a result, this M-strings set up becomes the $(p,q)$-fivebrane web in the type I\hspace{-.1em}IB theory (Fig.\ref{equivalence}(b) for $M = 2$ and $N = 2$). Finally, by deforming the theory by introducing the mass, this brane web is dual to the topological string theory on the non-compact toric Calabi-Yau manifold. Note that this mass deformation does not break supersymmetry any more. For this Calabi-Yau manifold, we can use the refined topological vertex formalism. By choosing the preferred direction appropriately, we can obtain the partition function of the M-string which is written in the form of a theta function.

On the other hand, by taking the $X_{7}$ direction as the M theory circle and performing T-duality along the $X_{6}$ direction\footnote{Even if we compactify the $X_{6}$ direction which we do not consider on the topological string theory side, this compactification does not affect the final result since the elliptic genus which we will calculate does not depend on the $X_{6}$ circle.},
we obtain another type I\hspace{-.1em}IB theory. $M$ M5-branes and M2-branes become the geometry of A$_{M - 1}$-type and D1-branes, respectively, and $\text{A}_{N-1}$ ALE becomes D5-branes. Therefore, the M-strings are dual to the theory on D1-branes that has been identified with a 2d $\Ncal = ( 0, 4 )$ U$(k)$ gauge theory \cite{Okuyama:2005gq}. In \cite{Haghighat:2013gba, Haghighat:2013tka}, for $M = 2$ and $N = 1$, they calculate the elliptic genus of the $\mathcal{N}=(0,4)$ gauge theory with an adjoint hypermultiplet $B$, a fundamental hypermultiplet $H$, and a Fermi multiplet $\La$, and show the equivalence between the refined topological string partition function and the elliptic genus. This equivalence means that we can obtain the partition function of the M-strings by calculating the partition function of the refined topological string.
We summarize the above duality chain in Fig.\ref{equivalence}.


\subsection{M-strings with the surface operator}
\begin{figure}[t]
\centering
    \includegraphics[width=4.5cm]{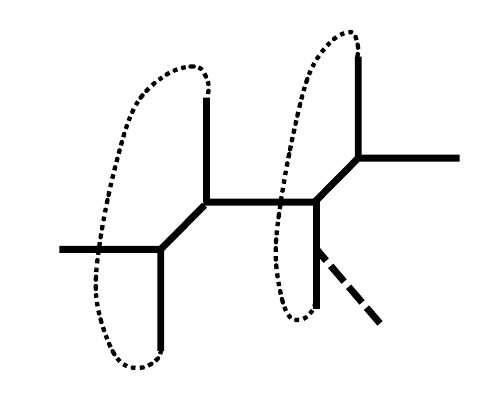}
    \caption{The web diagram with a Lagrangian brane.}
    \label{websurface}
\end{figure}
The surface operator in the gauge theory can be engineered from purely geometric languages with which the topological string theory is available \cite{Dimofte:2010tz, Taki:2010bj}. From now on, we consider one simple comprehensive example with inserting the Lagrange submanifold in Fig.\ref{websurface}. According to \cite{Dimofte:2010tz, Taki:2010bj}, this web diagram is dual to the D5-D3-NS5 system in type I\hspace{-.1em}IB theory, where the Lagrange submanifold corresponds to the D3-brane. Since the surface operator in the $\Omega$-background must lie in the space $\mathbb{R}^2 \subset \mathbb{C}^2$, the brane configuration is as in Table \ref{IIB}.
\par
\begin{table}[htb]
\centering
    \begin{tabular}{|c||c|c|c|c|c|c|c|c|c|c|} \hline
     IIB 
     & $X_{0}$ &  $X_{2}$ & $X_{3}$ & $X_{4}$ & $X_{5}$ & $X_{6}$ & $X_{7}$ 
     & $X_{8}$ & $X_{9}$ & $X_{10}$  \\ \hline \hline
    $2$ D5 
    &$\circ$&$\circ$&$\circ$&$\circ$&$\circ$& &$\circ$& & & \\ \hline
    $1$ NS5
    &$\circ$&$\circ$&$\circ$&$\circ$&$\circ$&$\circ$&  &  & & \\ \hline
    $k$ F1
    &$\circ$& & & & &$\circ$& & & & \\ \hline
    $1$ D3
    &$\circ$&$\circ$&$\circ$& & & & &$\circ$& & \\ \hline
    
  \end{tabular}
  \caption{The brane configuration which corresponds to the gauge theory with the surface operator. Since $X_{0}$ direction is compactified, the D3-brane can be interpreted as the surface operator in the gauge theory.}
    \label{IIB}
  \end{table}

\begin{figure}[htb]
\centering
    \includegraphics[width=4cm]{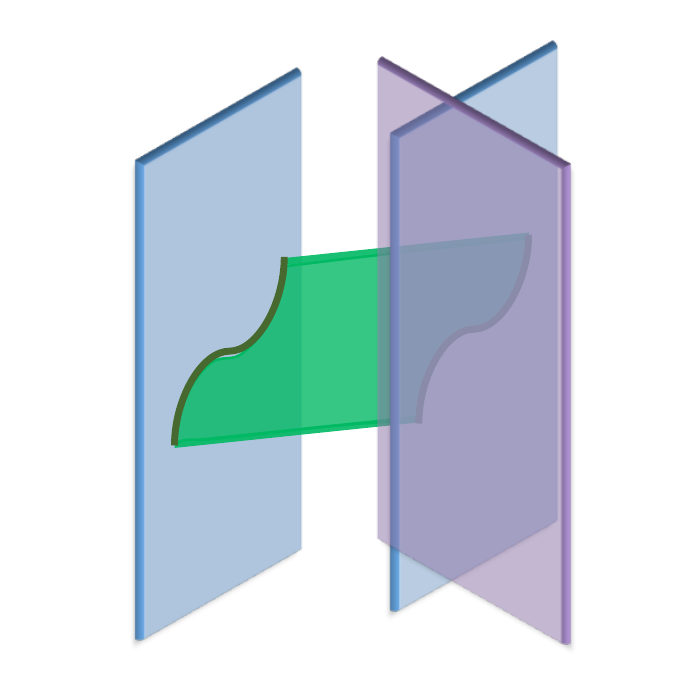}\vspace{1.5em}
  \begin{tabular}{|c||c|c|c|c|c|c|c|c|c|c|c|} \hline
     M-theory
     & $X_{0}$ & $X_{1}$ & $X_{2}$ & $X_{3}$ & $X_{4}$ & $X_{5}$ & $X_{6}$ & $X_{7}$ 
     &$X_{8}$ & $X_{9}$ & $X_{10}$  \\ \hline \hline
    $2$ M5
    &$\circ$ & $\circ$ & $\circ$ & $\circ$ & $\circ$ & $\circ$  &  &  &  & & \\ \hline
    $k$ M2
    &$\circ$ & $\circ$ &  &  &  &  & $\circ$ &  &  & & \\ \hline
    $1$ M5'
    &$\circ$&$\circ$&$\circ$&$\circ$& & & &$\circ$&$\circ$& & \\ \hline
    A$_{0}$  ALE
    & & & & & & & &$\circ$&$\circ$&$\circ$&$\circ$ \\ \hline
  \end{tabular}
  \caption
  {
  The M-theory brane system including the surface operator. The surface operator becomes a single M5-brane (M5' in the table). Although the A$_{0}$ ALE space is flat, we describe this space in order to connect easily to the type IIB theory.
  }
  \label{M-stringswithsurface}
 \end{figure}

Then, by performing T-duality along the $X_{7}$ directions, the D5-branes change D4-branes, the NS5-brane becomes $A_{0}$ ALE which is an almost flat space but with the point where the 1-cycle shrinks, and the D3-brane becomes the D4-brane', respectively. Finally, by lifting up to M-theory, we obtain the M-theory brane system. Thus, we conclude that the surface operator is dual to the M5-brane' whose extending direction is different from the one of the original M5-branes as we show in Fig.\ref{M-stringswithsurface}.

\begin{figure}[t] 
	\begin{center}
	\includegraphics[width=13cm,bb= 0 0 410 180,clip]{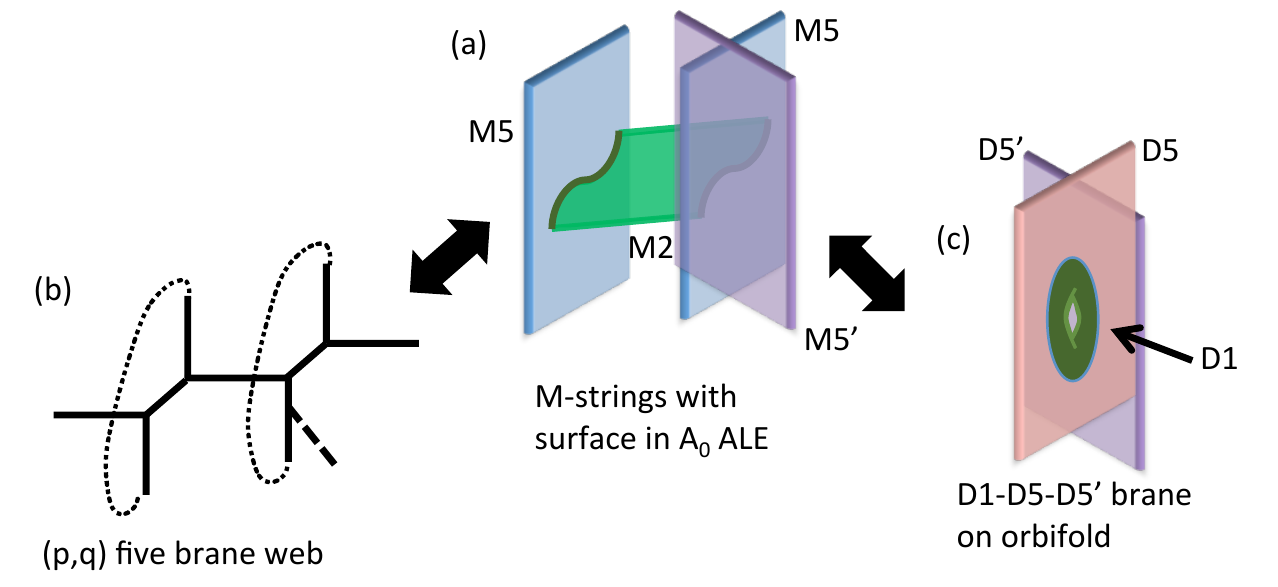}
	\caption{The duality chain as the open topological string extension. The purple sheet in figure (a) is an additional M5-brane (M5') which is dual to the D5-brane (D5') in figure (c).}
	\label{equivalence2}
	\end{center}
\end{figure}

As is the case of the duality chain in the previous subsection, let us take the $X_{7}$ direction as the M-theory circle and perform T-duality along the $X_{6}$ direction.
The net effect reduces the M-strings Fig.\ref{M-stringswithsurface} to the D1-D5 system on A$_1$ ALE with an additional D5-brane' wrapped on the $012368$ directions. We should note that the $68$ directions are the subspace of A$_1$ ALE.
We summarize the duality chain in Fig.\ref{equivalence2} which is analogous to the previous one.
\par
Let us consider how many supercharges the M-strings have. The conditions under which the supercharges are conserved are given by
\begin{eqnarray}
\Gamma^{0}\Gamma^{1}\Gamma^{2}\Gamma^{3}\Gamma^{4}\Gamma^{5}\epsilon
=
\Gamma^{0}\Gamma^{1}\Gamma^{6}\epsilon
=
\Gamma^{0}\Gamma^{1}\Gamma^{2}\Gamma^{3}\Gamma^{7}\Gamma^{8}\epsilon
=\epsilon,
\end{eqnarray}
where $\epsilon$ is the 32-component Killing spinor, and $\Gamma^{I}$ is the Gamma matrices in 11 dimensions. Then, we can show that there are 4 preserved supercharges. Since there is no reason to be a theory with chiral supersymmetry, we can expect that our M-strings have $\mathcal{N}=(2,2)$ supersymmetry. However, because of the mass deformation, the supersymmetry is broken to $\mathcal{N}=(0,2)$.

\subsection{Calculations of the partition function in the M-strings}
\begin{figure}[t]
\centering
    \includegraphics[width=11.5cm,bb= 270 240 570 360,clip]{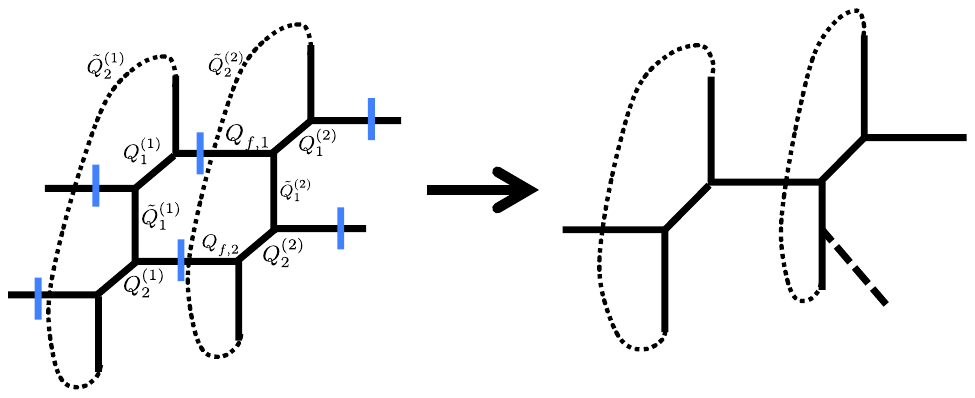}
    \caption{The assignment of the K\"ahler parameters on the web diagram for $M=N=2$ which engineers the fundamental surface operator in the degenerate limit.} 
    \label{AGT}
\end{figure}

In this subsection, we calculate the partition function of the M-strings with the surface operator. In order to do it, we have to consider the web diagram of Fig.\ref{websurface}.  However, the calculation of this web diagram is rather complicated. Instead, we propose an alternative method of calculation. The key ideas are AGT correspondence in the presence of the surface operator \cite{Alday:2009fs} and the bubbling Calabi-Yau \cite{Taki:2010bj} shown in the previous section. As explained in the previous section, the fact that the insertion of the surface operator corresponds to that of the degeneration field in AGT correspondence is encoded into the topological string formalism (Fig.\ref{AGT}) through the limit of the K\"ahler parameters corresponding to \eqref{limK}. We shown that both diagrams in Fig.\ref{AGT} have the M-theory origins, which means that we can obtain the partition function of the M-strings with the surface operator from the refined topological string.
\par
In order to calculate the partition function in the presence of the surface operator, we start the web diagram of the left figure of Fig.\ref{AGT} with defining the K\"ahler parameters for all intervals.
\begin{figure}[t]
\centering
    \includegraphics[width=4.5cm,bb= 200 325 390 540,clip]{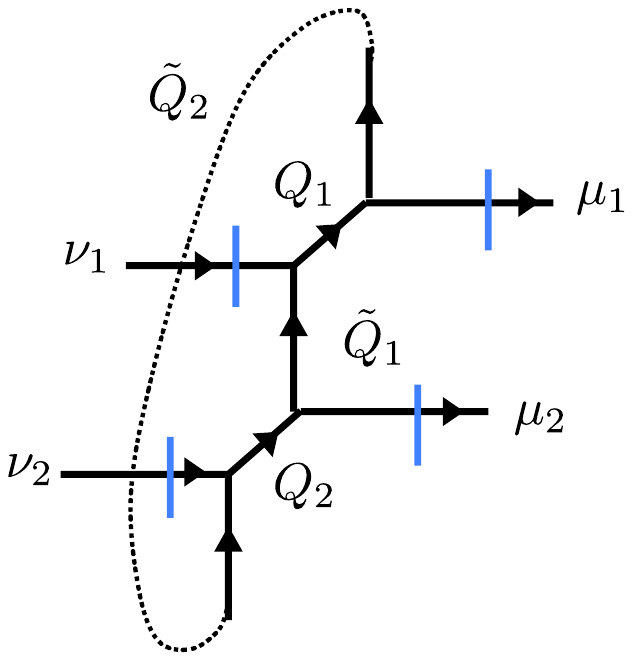}
    \caption{The building block to compute the partition function. The preferred direction is taken to be horizontal (blue).}
    \label{webdiagram}
\end{figure}
We take the preferred direction to be horizontal, and we can write down the partition function by using the topological vertex formalism,
\begin{eqnarray}
\mathcal{Z}_{\text{top}} (Q;t,q)
=
\sum_{\mu_{1},\mu_{2}}(-Q_{f,1})^{|\mu_{1}|}(-Q_{f,2})^{|\mu_{2}|}
\mathcal{Z}_{\mu_{1}\mu_{2}\emptyset\emptyset}^{\text{build}}(Q^{(1)};t,q)
\mathcal{Z}_{\emptyset\emptyset\mu_{1}\mu_{2}}^{\text{build}}(Q^{(2)};t,q),
\end{eqnarray}
where we define the building block $\mathcal{Z}_{\mu_{1}\mu_{2}\nu_{1}\nu_{2}}^{\text{build}}(Q;t,q)$ (Fig.\ref{webdiagram}) as
\begin{eqnarray}
\mathcal{Z}^{\text{build}}_{\mu_{1}\mu_{2}\nu_{1}\nu_{2}}(Q;t,q)
&=&
\sum_{\eta_{1,2},\tilde{\eta}_{1,2}}
(-Q_{1})^{|\eta_{1}|}(-Q_{2})^{|\eta_{2}|}(-\tilde{Q}_{1})^{|\tilde{\eta}_{1}|}(-\tilde{Q}_{2})^{|\tilde{\eta}_{2}|}
\nonumber \\ &&\times
C_{\tilde{\eta}_{2}\eta_{1}^{t}\mu_{1}}(t,q)C_{\tilde{\eta}_{1}^{t}\eta_{1}\nu_{1}^{t}}(q,t)
C_{\tilde{\eta}_{1}\eta_{2}^{t}\mu_{2}}(t,q)C_{\tilde{\eta}_{2}^{t}\eta_{2}\nu_{2}^{t}}(q,t),
\end{eqnarray}
with the refined topological vertex $C_{\la \mu \nu} ( t, q )$ defined in Appendix \ref{Convension}, and the variables $t,~q$ are defined as
\begin{eqnarray}
t=e^{2\pi \mathrm{i}\epsilon_{1}},~q=e^{-2\pi \mathrm{i}\epsilon_{2}}.
\end{eqnarray}
After some calculation, we obtain \cite{Haghighat:2013tka}
\begin{align}
\hat{\mathcal{Z}}_{\text{top}} (Q;t,q)
&:=
\frac
{\mathcal{Z}_{\text{top}} (Q;t,q)}
{\mathcal{Z}_{\text{top}} (Q;t,q)|_{Q_{f,1},Q_{f,2}=0}}
\nonumber \\
&=
\sum_{\mu_{1},\mu_{2}}
\biggl(-Q_{f,1}\sqrt{\frac{q}{t}}Q_{1}^{(1)}Q_{2}^{(1)}\biggr)^{|\mu_{1}|}
\biggl(-Q_{f,2}\sqrt{\frac{t}{q}}Q_{1}^{(2)}Q_{2}^{(2)}\biggr)^{|\mu_{2}|}
\nonumber \\
&\times~~
\prod_{(i,j)\in\mu_{1}}
\Biggl[
\frac{
\theta_{1}(\tau ; Q_{1}^{(2)^{-1}}t^{-\mu_{1,i}+j-\frac{1}{2}}q^{i-\frac{1}{2}})
\theta_{1}(\tau ;  Q_{1}^{(1)^{-1}}t^{\mu_{1,i}-j+\frac{1}{2}}q^{-i+\frac{1}{2}})
}{
\theta_{1}(\tau ; t^{\mu_{1,i}-j}q^{\mu^{t}_{1,j}-i+1})
\theta_{1}(\tau ; t^{\mu_{1,i}-j+1}q^{\mu^{t}_{1,j}-i})} 
 \nonumber \\
 &\times~~
\frac{
\theta_{1}(\tau ; Q_{1}^{(2)^{-1}}Q_{2}^{(2)^{-1}}\tilde{Q}_{2}^{(2)^{-1}}t^{-\mu_{1,i}+j-\frac{1}{2}}q^{i-\frac{1}{2}})
\theta_{1}(\tau ;  Q_{2}^{(1)^{-1}}Q_{1}^{(1)^{-1}}\tilde{Q}_{1}^{(1)^{-1}}t^{\mu_{1,i}-j+\frac{1}{2}}q^{-i+\frac{1}{2}})
}{
\theta_{1}(\tau ; Q_{2}^{(1)^{-1}}\tilde{Q}_{2}^{(1)^{-1}}t^{-\mu_{1,i}+j-1}q^{-\mu^{t}_{2,j}+i})
\theta_{1}(\tau ; Q_{1}^{(1)^{-1}}\tilde{Q}_{1}^{(1)^{-1}}t^{\mu_{1,i}-j}q^{\mu^{t}_{2,j}-i+1})} \Biggr]
 \nonumber \\
&\times~~
\prod_{(i,j)\in\mu_{2}}\Biggl[
\frac{
\theta_{1}(\tau ; Q_{2}^{(2)^{-1}}t^{-\mu_{2,i}+j-\frac{1}{2}}q^{i-\frac{1}{2}})
\theta_{1}(\tau ;  Q_{2}^{(1)^{-1}}t^{\mu_{2,i}-j+\frac{1}{2}}q^{-i+\frac{1}{2}})
}{
\theta_{1}(\tau ; t^{\mu_{2,i}-j}q^{\mu^{t}_{2,j}-i+1})
\theta_{1}(\tau ; t^{\mu_{2,i}-j+1}q^{\mu^{t}_{2,j}-i})
} 
\nonumber \\
&\times~~
\frac{
\theta_{1}(\tau ; Q_{1}^{(1)^{-1}}Q_{2}^{(1)^{-1}}\tilde{Q}_{2}^{(1)^{-1}}t^{\mu_{2,i}-j+\frac{1}{2}}q^{-i+\frac{1}{2}})
\theta_{1}(\tau ;  Q_{2}^{(2)^{-1}}Q_{1}^{(2)^{-1}}\tilde{Q}_{1}^{(2)^{-1}}t^{-\mu_{2,i}+j-\frac{1}{2}}q^{i-\frac{1}{2}})
}{
\theta_{1}(\tau ; Q_{2}^{(1)^{-1}}\tilde{Q}_{2}^{(1)^{-1}}t^{\mu_{2,i}-j}q^{\mu^{t}_{1,j}-i+1})
\theta_{1}(\tau ; Q_{1}^{(1)^{-1}}\tilde{Q}_{1}^{(1)^{-1}}t^{-\mu_{2,i}+j-1}q^{-\mu^{t}_{1,j}+i})
}
 \Biggr], \nonumber \\
\label{top. str. part. fn.}
\end{align}
where we define the parameter $Q_{\tau}=Q^{(1)}_{1}\tilde{Q}^{(1)}_{1}Q^{(1)}_{2}\tilde{Q}^{(1)}_{2}=Q^{(2)}_{1}\tilde{Q}^{(2)}_{1}Q^{(2)}_{2}\tilde{Q}^{(2)}_{2}=e^{2\pi\mathrm{i}\tau}$.
\par
Now, we consider the insertion of the surface operator. According to AGT correspondence, we set the K\"ahler parameters as \eqref{limK}, that is,
\begin{eqnarray}
Q_{2}^{(1)}=\sqrt\frac{t}{q},~~Q_{2}^{(2)}=\frac{1}{q}\sqrt{\frac{t}{q}}.
\end{eqnarray}
Then, the products of (\ref{top. str. part. fn.}) contain the factor $\prod_{(i,j)\in\mu_{2}}\theta_{1}(\tau ;  t^{\mu_{2,i}-j}q^{-i+1})$. Unless the Young diagram $\mu_{2}$ is empty, this factor become zero due to $\theta_{1}(\tau ;  1)=0$. Therefore, we need to set $\mu_{2}=\emptyset$. Then, after several cancellations, we obtain
\begin{eqnarray}
\hat{\mathcal{Z}}_{\text{top}} (Q;t,q)
 &=&
 \sum_{\mu_{1}}\lp-Q_{f,1} Q_{m}\rp^{|\mu_{1}|}
 \nonumber \\ &&\times~~
\prod_{(i,j)\in\mu_{1}}
\Biggl[
\frac{
\theta_{1}(\tau; Q_{m}^{-1}t^{-\mu_{1,i}+j-\frac{1}{2}}q^{i-\frac{1}{2}})
\theta_{1}(\tau ;  Q_{m}^{-1}t^{\mu_{1,i}-j+\frac{1}{2}}q^{-i+\frac{1}{2}})
}{
\theta_{1}(\tau ; t^{\mu_{1,i}-j}q^{\mu^{t}_{1,j}-i+1})
\theta_{1}(\tau ; t^{\mu_{1,i}-j+1}q^{\mu^{t}_{1,j}-i})}
\Biggr]
 \nonumber \\ &&\times~~
\prod_{(i,j)\in\mu_{1}}
\frac{
\theta_{1}(\tau ; Q_{m}^{-1}\tilde{Q}_{2}^{(2)^{-1}}t^{-\mu_{1,i}+j-1}q^{i+1})
}{
\theta_{1}(\tau; Q_{m}^{-1}\tilde{Q}_{2}^{(2)^{-1}}t^{-\mu_{1,i}+j-1}q^{i})
}
\label{generator},
\end{eqnarray}
where we set $Q_{1}^{(1)}=Q_{1}^{(2)}=:Q_{m}$. Thus, we obtain the partition function of $k$ M-strings with the surface operator
\begin{eqnarray}
\hat{\mathcal{Z}}_{\text{M-strings}}^{(k)} (Q;t,q)
 &=&
 \sum_{|\mu_{1}|=k}
\prod_{(i,j)\in\mu_{1}}
\Biggl[
\frac{
\theta_{1}(\tau ; Q_{m}^{-1}t^{-\mu_{1,i}+j-\frac{1}{2}}q^{i-\frac{1}{2}})
\theta_{1}(\tau ;  Q_{m}^{-1}t^{\mu_{1,i}-j+\frac{1}{2}}q^{-i+\frac{1}{2}})
}{
\theta_{1}(\tau; t^{\mu_{1,i}-j}q^{\mu^{t}_{1,j}-i+1})
\theta_{1}(\tau ; t^{\mu_{1,i}-j+1}q^{\mu^{t}_{1,j}-i})}
\Biggr]
 \nonumber \\ &&\times~~
\prod_{(i,j)\in\mu_{1}}
\frac{
\theta_{1}(\tau; Q_{m}^{-1}\tilde{Q}_{2}^{(2)^{-1}}t^{-\mu_{1,i}+j-1}q^{i+1})
}{
\theta_{1}(\tau ; Q_{m}^{-1}\tilde{Q}_{2}^{(2)^{-1}}t^{-\mu_{1,i}+j-1}q^{i})
}
\label{k M-string}
.
\end{eqnarray}
Note that the first product of $\mu_{1}$ are the same as the partition function of the M-strings for $( M, N ) = ( 2, 1 )$ without the surface operator. Thus, we can expect that the second product is the contribution from the surface operator.

\subsection{Comparison with the $\Ncal=(2,2)$ elliptic genus}
In this subsection, we consider the 2d gauge theory description which corresponds to our M-strings (Fig.\ref{M-stringswithsurface}).
As commented, the M-strings on the top of the singularity without the surface operator preserves $( 0, 4 )$ supersymmetry, and the 2d theory dual to the M-strings with $( M, N ) = ( 2, 1 )$ is U$(k)$ gauge theory with $\Ncal = ( 0, 4 )$ multiplets $( B, H, \La )$. Although those matter contents can be found directly by the analysis of open string spectrum in the D1-D5 on the ALE space shown in Fig.\ref{equivalence}(c) \cite{Okuyama:2005gq}, for avoiding complexity from orbifold to identify them, we move to another dual frame without a singular background which where the same matters as in the D1-D5 on the ALE space can be read. When thinking of the M-theory circle as the $X_7$ direction and implementing T-duality along the $X_1$ direction, the standard M-strings of Fig.\ref{equivalence}(a) is translated into the following brane system \cite{Haghighat:2013tka}:
\begin{align} 
    \begin{tabular}{|c||c|c|c|c|c|c|c|c|c|c|} \hline
    I\hspace{-.1em}IB & $X_0$ & $X_1$ & $X_2$ & $X_3$ & $X_4$ & $X_5$ & $X_6$ & $X_8$ & $X_9$ & $X_{10}$ \\ \hline \hline 
    $2$ NS5 & $\circ$ & $\circ$ & $\circ$ & $\circ$ & $\circ$ & $\circ$ & $\{ a_i \}$ &  &  &  \\ \hline 
    $k$ D1 & $\circ$ &  &  &  &  &  & $\circ$ &  &  &  \\ \hline 
    $1$ D5 & $\circ$ &  & $\circ$ & $\circ$ & $\circ$ & $\circ$ & $\circ$ &  &  &  \\ \hline 
    \end{tabular}
    \label{gaugeIIBa}
\end{align}
where $a_i$ are positions of NS5-branes in the $X_6$ direction. The worldvolume theory of D1-branes in this dual frame produces the same gauge theory mentioned above.

Introducing the half-BPS surface operator breaks $( 0, 4 )$ supersymmetry down to $( 0, 2 )$. We would like to identify the 2d $( 0, 2 )$ theory which can describe the M-strings with the elementary surface operator. In the dual I\hspace{-.1em}IB picture shown in Fig.\ref{equivalence2}(c), It is not so simple to read off the matters of the 2d gauge theory on the D1-branes since the D5-brane' mapped from the M5-brane' is wrapped on the two-dimensional subspace of A$_1$ ALE originated from two M5-branes. To guess the $( 0, 2 )$ multiplets, we honestly follow the prescription to generate the brane system \eqref{gaugeIIBa}. As a result, the M-strings in the presence of M5' has the following dual frame:
\begin{align} 
    \begin{tabular}{|c||c|c|c|c|c|c|c|c|c|c|} \hline
    I\hspace{-.1em}IB & $X_0$ & $X_1$ & $X_2$ & $X_3$ & $X_4$ & $X_5$ & $X_6$ & $X_8$ & $X_9$ & $X_{10}$ \\ \hline \hline 
    $2$ NS5 & $\circ$ & $\circ$ & $\circ$ & $\circ$ & $\circ$ & $\circ$ & $\{ a_i \}$ &  &  &  \\ \hline 
    $k$ D1 & $\circ$ &  &  &  &  &  & $\circ$ &  &  &  \\ \hline 
    $1$ D3' & $\circ$ &  & $\circ$ & $\circ$ &  &  &  & $\circ$ &  &  \\ \hline 
    $1$ D5 & $\circ$ &  & $\circ$ & $\circ$ & $\circ$ & $\circ$ & $\circ$ &  &  &  \\ \hline 
    \end{tabular}
    \label{gaugeIIBpa}
\end{align}
Inserting the surface operator leads to the appearance of an D3-brane'. In this frame, in addition to $\Ncal = ( 0, 4 )$ multiplets $( B, H, \La )$, there might be $( 0, 2 )$ chiral $\tilde{\phi}$ and Fermi multiplets $\tilde{\psi}$ coming from open strings ending on D1 and D3'. It is well-known that the $( 0, 4 )$ multiplets are decomposed into the $( 0, 2 )$ ones,
\begin{align} 
	\begin{aligned}
	( 0, 4 ) \text{ vector} &\to ( 0, 2 ) \text{ vector} + \text{one } ( 0, 2 ) \text{ adjoint Fermi } \tilde{\xi}, \\ 
	( 0, 4 ) \text{ adjoint hyper } B &\to \text{two } ( 0, 2 ) \text{ adjoint chirals } \tilde{b}_{1, 2}, \\ 
	( 0, 4 ) \text{ fundamental hyper } H &\to \text{two } ( 0, 2 ) \text{ fundamental chirals } \tilde{h}_{1, 2}, \\ 
	( 0, 4 ) \text{ Fermi } \La &\to \text{two } ( 0, 2 ) \text{ Fermis } \tilde{\la}_{1, 2}. 
	\end{aligned}
\end{align}
Namely, we propose that the M-strings with the fundamental surface operator can be given by the 2d $( 0, 2 )$ U$(k)$ gauge theory with adjoint chiral multiplets $\tilde{b}_{1, 2}$, an adjoint Fermi $\tilde{\xi}$, fundamental chiral multiplets $( \tilde{h}_{1, 2}, \tilde{\phi} )$, and Fermi multiplets $( \tilde{\la}_{1, 2}, \tilde{\psi} )$.

In the rest here, we aim to compute the elliptic genus of the 2d gauge theory and compare it with the partition function of our M-strings \eqref{k M-string} as a consistency check of our expectation. The complete formulae of the $( 0, 2 )$ elliptic genera have been derived in \cite{Benini:2013nda, Benini:2013xpa} by the localization technique. Since the elliptic genera generically have many poles, we must be carefully choose the contour around the poles in the integral over Coulomb parameters, i.e., expectation values of the scalar in the vector multiplet. The standard method in completing the integral with choosing appropriate paths is called the Jeffrey-Kirwan residue.
However, this calculation seems quite intricate even in our situation with the elementary surface operator. Instead, we adopt Hosomichi-Lee's method \cite{Hosomichi:2014rqa} baked on the Higgs branch localization, in which we need not perform the contour integral after the localization. Moreover, since the result does not depend on the choice of the localization method, we can use the Higgs branch localization method.

Before showing the details of the elliptic genus, in order to follow the way of \cite{Hosomichi:2014rqa}, we recall that $( 0, 4 )$ supersymmetry on the M-strings without the surface operator gets enhanced to $( 4, 4 )$ \cite{Haghighat:2013gba} when we set $Q_{m} = \sqrt{t/q}$ (equivalently, $2 m = \e_1 + \e_2$). With this specialization in our case, supersymmetry may be enlarged to $( 2, 2 )$ because tuning the mass is not affected by inserting the surface operator. In what follows, we focus on checking if this expectation is correct.
For this purpose, with formulae,
\begin{align} 
	\begin{aligned}
	\theta_{1}(\tau|z+\tau)
	&=
	-e^{-2\pi \mathrm{i} z-\mathrm{i}\pi \tau}\theta_{1}(\tau| z) \\ 
	\theta_{1}(\tau|-z)
	&=
	-\theta_{1}(\tau|z), \\ 
	\prod_{(i,j)\in \nu}(1-Qq^{j-\frac{1}{2}}t^{-i+\frac{1}{2}})
	&=
	\prod_{(i,j)\in \nu}(1-Qq^{\nu_{i}-j+\frac{1}{2}}t^{-i+\frac{1}{2}}), 
	\end{aligned}
\end{align}
and the relation $Q_{\tau} = \frac{1}{q}\sqrt{\frac{t}{q}} Q_{1}^{(2)}\tilde{Q}_{1}^{(2)}\tilde{Q}_{2}^{(2)}$, we can rewrite (\ref{generator}) as
\begin{align}
\hat{\mathcal{Z}}_{\text{M-strings}}^{( k )} (Q;t,q)
&=
\sum_{|\mu_{1}|=k}
\prod_{(i,j)\in\mu_{1}}
\Biggl[
\frac{
\theta_{1}(\tau; Q_{m}^{-1}t^{-j+\frac{1}{2}}q^{i-\frac{1}{2}})
\theta_{1}(\tau ;  Q_{m}^{-1}t^{j-\frac{1}{2}}q^{-i+\frac{1}{2}})
}{
\theta_{1}(\tau ; t^{\mu_{1,i}-j}q^{\mu^{t}_{1,j}-i+1})
\theta_{1}(\tau ; t^{\mu_{1,i}-j+1}q^{\mu^{t}_{1,j}-i})}
\Biggr]
\nonumber \\
&\hspace{1em} \times~~
\prod_{(i,j)\in\mu_{1}}
\frac{
\theta_{1}(\tau ; q\tilde{Q}_{1}^{(2)^{-1}}t^{j-\frac{1}{2}}q^{-i-\frac{1}{2}})
}{
\theta_{1}(\tau; q\tilde{Q}_{1}^{(2)^{-1}}t^{j-\frac{1}{2}}q^{-i+\frac{1}{2}})
}.
\label{M-stsurface}
\end{align}
Then, we need to set $Q_{m}=\sqrt{t/q}$ to see the supersymmetry enhancement. Note that substituting this value into (\ref{M-stsurface}) naively results in zero due to the fermion zero modes.
Therefore, we should divide (\ref{M-stsurface}) by $\hat{\mathcal{Z}}_{\text{M-strings}}^{( 1 )} (Q;t,q)$. Then, we obtain
\begin{align}
\frac{
\hat{\mathcal{Z}}_{\text{M-strings}}^{( k )} (Q;t,q)
}{\hat{\mathcal{Z}}_{\text{M-strings}}^{( 1 )} (Q;t,q)
}
&=
\sum_{|\mu_{1}|=k}
\Biggl[
\frac{
\prod_{(i,j)\in \mu,(i,j)\neq(1,1)}
\theta_{1}(\tau ; t^{-j}q^{i})
\theta_{1}(\tau ; t^{j-1}q^{-i+1})
}{
\lc
\prod_{(i,j)\in \mu}
\theta_{1}(\tau ; t^{\mu_{1,i}-j}q^{\mu^{t}_{1,j}-i+1})
\theta_{1}(\tau ; t^{\mu_{1,i}-j+1}q^{\mu^{t}_{1,j}-i})
\rc
/
\lc
\theta_{1}(\tau;t)
\theta_{1}(\tau;q)
\rc
}
\Biggr]
\nonumber \\ 
&\hspace{1em} \times~~
\frac{
\prod_{(i,j)\in \mu,(i,j)\neq(1,1)}
\theta_{1}(\tau ; q\tilde{Q}_{1}^{(2)^{-1}}t^{j-\frac{1}{2}}q^{-i+\frac{1}{2}})
}{
\prod_{(i,j)\in \mu,(i,j)\neq(1,1)}
\theta_{1}(\tau ; q\tilde{Q}_{1}^{(2)^{-1}}t^{j-\frac{1}{2}}q^{-i+\frac{3}{2}})
}. 
\label{44Mstring}
\end{align}
Indeed, the first line of \eqref{44Mstring} is completely identical to the partition function without the insertion of M5' derived by \cite{Haghighat:2013gba}. Namely, the additional effect coming from M5' is expressed as the second line of \eqref{44Mstring}.

Let us turn to the 2d field theory description of the M-strings. 
As explained from the viewpoint of the D1-D5-NS5 system with D3' \eqref{gaugeIIBpa}, for general $Q_m$, this keeps $( 0, 2 )$ supersymmetry with the U$(k)$ gauge symmetry and the matter contents shown above. These may nicely be combined into $( 2, 2 )$ multiplet, but we further anticipate cancelling the contributions of one chiral and one Fermi in the $( 0, 2 )$ one-loop determinant as $Q_m = \sqrt{t/q}$. This is because in this limit the open strings with the end at D5 and D3' neighboring on D1 can become massless.
Then, by using the localization formula \cite{Benini:2013nda, Benini:2013xpa}, we can write down the one-loop determinant of these matter contents,
\begin{align} 
\mathcal{Z}_{\text{1-loop}}
&=
\Biggl( 
\frac{
2 \pi \eta ( \tau )^3
}{
\theta_{1}(\tau|\epsilon_{2})
}
\Biggr)^k
\Biggl( 
\prod_{i \neq j}^k
\frac{
\theta_{1}(\tau|u_{i}-u_{j})
}{
\theta_{1}(\tau|u_{i}-u_{j}+\epsilon_{2})
}
\Biggr)
\Biggl( 
\prod_{i,j=1}^k
\frac{
\theta_{1}(\tau|u_{i}-u_{j}+\epsilon_{1}+\epsilon_{2})
}{
\theta_{1}(\tau|u_{i}-u_{j}+\epsilon_{1})
}
\Biggr)
\nonumber \\
&\hspace{1em} \times
\Biggl( 
\prod_{i=1}^k
\frac{
\theta_{1}(\tau|u_{i}+\chi+\epsilon_{2})
}{
\theta_{1}(\tau|u_{i}+\chi)
}
\Biggr),
\label{oneloop}
\end{align}
where the variables $u_{i},~\tau,~\text{and}~\chi$ are the integration variables, a complex structure modulus of the torus, and the chemical potential associated to a flavor symmetry, respectively. Notice that the $\Ncal = ( 2, 2 )$ vector multiplet and the adjoint chiral multiplet put on the first line of \eqref{oneloop} are brought together into the $\Ncal = ( 4, 4 )$ vector, and it is found in \cite{Hosomichi:2014rqa} that its elliptic genus coincides with the M-strings partition function shown in the first line of \eqref{44Mstring}. Because these observations are consist with the result in \cite{Hosomichi:2014rqa}, we can follow them to re-express the elliptic genus with the one-loop determinant \eqref{oneloop}:
The vacua in the Higgs branch localization are satisfied with the D-term and F-term conditions from equations of motion for these auxiliary fields and BPS equations and finally are labelled by a Young tableau. We propose the following formula for the elliptic genus based on their arguments:
%
\begin{align} 
\mathcal{Z}^{( k )}_{\text{Ell}}
=
\sum_{\{ u_{i} \}}
\prod_{i,j=1}^{k}
\frac{
\theta_{1}(\tau|u_{i}-u_{j})
\theta_{1}(\tau|u_{i}-u_{j}+\epsilon_{1}+\epsilon_{2})
}{
\theta_{1}(\tau|u_{i}-u_{j}+\epsilon_{1})
\theta_{1}(\tau|u_{i}-u_{j}+\epsilon_{2})
}
\times
\prod_{i=1}^{k}
\frac{
\theta_{1}(\tau|u_{i}+\chi+\epsilon_{2})
}{
\theta_{1}(\tau|u_{i}+\chi)
},
\end{align}
where the sum is taken over all the fixed points in the Higgs branch. Combining the BPS conditions with the D-term and F-term conditions, the set of the fixed points as the solutions to these conditions is parametrized by a Young tableau with $k$ boxes, 
\begin{align} 
(\bold{u})_{i,j}
=
\Bigl(j+\frac{1}{2}\Bigr)\epsilon_{1} + \Bigl(i+\frac{1}{2}\Bigr)\epsilon_{2},
\end{align}
where we define $\bold{u}:=(u_{1},u_{2},...,u_{k})$ and the indices $( i, j )$ represent a position in the Young tableau (see Appendix \ref{Convension} for its definition). 
Then the elliptic genus becomes
\begin{align} %
\frac{\mathcal{Z}^{( k )}_{\text{Ell}}}{\mathcal{Z}^{( 1 )}_{\text{Ell}}}
&=
\sum_{|\mu|=k}
\frac{
\prod_{(i,j)\in \mu,(i,j)\neq(1,1)}
\theta_{1}(\tau|(j-1)\epsilon_{1} + (i-1)\epsilon_{2})
\theta_{1}(\tau|-j\epsilon_{1}-i\epsilon_{2})
}{
(
\prod_{(i,j)\in \mu}
\theta_{1}(\tau|(j-\mu_{i})\epsilon_{1} + (\mu^{t}_{j}-i+1)\epsilon_{2})
\theta_{1}(\tau|(\mu_{i}-j+1)\epsilon_{1} + (i-\mu^{t}_{j})\epsilon_{2})
)
/
(
\theta_{1}(\tau|\epsilon_{1})
\theta_{1}(\tau|\epsilon_{2})
)
}
\nonumber \\
&\hspace{1em} \times
\frac{
\prod_{(i,j)\in \mu,(i,j)\neq(1,1)}
\theta_{1}(\tau|\chi+(j+\frac{1}{2})\epsilon_{1} + (i+\frac{3}{2})\epsilon_{2})
}{
\prod_{(i,j)\in \mu,(i,j)\neq(1,1)}
\theta_{1}(\tau|(\chi+(j+\frac{1}{2})\epsilon_{1} + (i+\frac{1}{2})\epsilon_{2})}
,
\label{ell}
\end{align}
where $\mathcal{Z}^{( 1 )}_{\text{Ell}}$ corresponds to the free U$(1)$ part of the U$(k)$ gauge group. The first line of \eqref{ell} is completely the same as eq.(5.24) in \cite{Hosomichi:2014rqa} (with setting $2 m = \e_1 + \e_2$ there), that is, the first line of \eqref{44Mstring}. Further, the second line of \eqref{ell} is perfectly agreement with that of \eqref{44Mstring} under the identification $t^{-1}q^2 \lp \tilde{Q}_{1}^{(2)} \rp^{-1}=e^{2\pi\mathrm{i}\chi}$. This agreement justifies our duality picture in the M-strings configuration which geometrically engineers the surface operator.

\section{Discussions}\label{Sum}
In summary, we found a new M-string configuration where an additional M5 intersecting with one of two parallel M5's can geometrically engineer the surface operator in the 4d gauge theory via AGT correspondence. Based on the string duality chain, this M-theory picture is translated into the standard web diagram corresponding to the open topological string. Therefore, the partition function of our M-strings under the twist can be evaluated by the refined topological vertex with the particular preferred direction. Further, taking T-duality in the another direction provides the D1-D5 system on the $\text{A}_1$ ALE space dual to our M-theory picture, which means that its partition function may also be identical to the elliptic genus of the 2d $\Ncal = ( 2, 2 )$ theory realized on D1's. We gave a nontrivial support for our proposal to the M-strings by seeing the agreement of the partition function of the refined topological string with the elliptic genus.

We would like to close the paper with some comments on interesting open questions. Firstly, our result for the new M-strings system must be mathematically equivalent to the 5d Nekrasov partition function derived in \cite{Taki:2010bj} since physics does not depend on the selection of the preferred direction of the refined topological vertex. It is necessary to establish this equality to be able to support our proposal on AGT correspondence. Further, the geometric transition technique which we accepted can be applied to more general bubbling Calabi-Yau's, namely, it seems possible to generate wide varieties of AGT correspondence in the presence of more general surface operators.

Secondly, although the bubbling geometry before the geometric transition is descended from M-theory on the $\text{A}_1$ ALE space, the toric brane diagram after the geometric transition can be uplifted to M-theory on the flat space but with the extra M5. There is no ambiguity in the meaning of the geometric transition from the standpoint of string theory, but it is obscure if the operation of the geometric transition connects two M-theory configurations on the absolutely distinct backgrounds. To answer how we can interpret it in the M-theory picture perhaps brings a new insight to M-theory.

Finally, we need to establish the direct check of preserving $\Ncal = ( 2, 2 )$ supersymmetry from the viewpoint of D1's on the $\text{A}_1$ ALE space. This amount of supersymmetry is basically expected from the M-strings with the insertion of the M5 on the flat space, and in fact $\Ncal = ( 2, 2 )$ have been observed in \cite{Hori:2013ewa} for the M2-branes suspended between a flat M5-brane and a curved M5-brane, which are actually similar to our M-strings. However, the dual picture of the D-D5 system includes the orbifolded background, that is, the $\text{A}_1$ ALE space. The way to put the branes on a orbifolded space is normally used to make supersymmetry be chiral as well as the original M-string configuration is dual to another D1-D5 system on the ALE space whose worldvolume theory keeps $\Ncal = ( 0, 4 )$ \cite{Okuyama:2005gq}, but this argument highly depends on which directions the background D5's extend to. We wonder whether there is a essential connection of the M5' in our M-strings to the curved one in \cite{Hori:2013ewa} and would like to explicitly clarify the appearance of $\Ncal = ( 2, 2 )$ supersymmetry. Those points become keys to verify our proposal. We hope that we can further report resolutions to the above issues in the near future.

\subsection*{Acknowledgment}
We would like to thank Kazuo Hosomichi, Taro Kimura, Yuji Okawa, Takuya Okuda, Masato Taki, and Satoshi Yamaguchi for useful discussions. We also would like to thank Babak Haghighat, Can Koz\c{c}az, and Wenbin Yan for telling us their progress which has some overlap with our calculations. The work of H.M. was supported in part by the JSPS Research Fellowship for Young Scientists.

\appendix
\section{Definition and notation}\label{Convension}

\subsection{Mathematical Definition}
\paragraph{Young diagram}
\par~\par
\noindent
We define the Young diagram $\mu$ as the following figure:
\begin{figure}[htbp]
\centering
    \includegraphics[width=15cm]{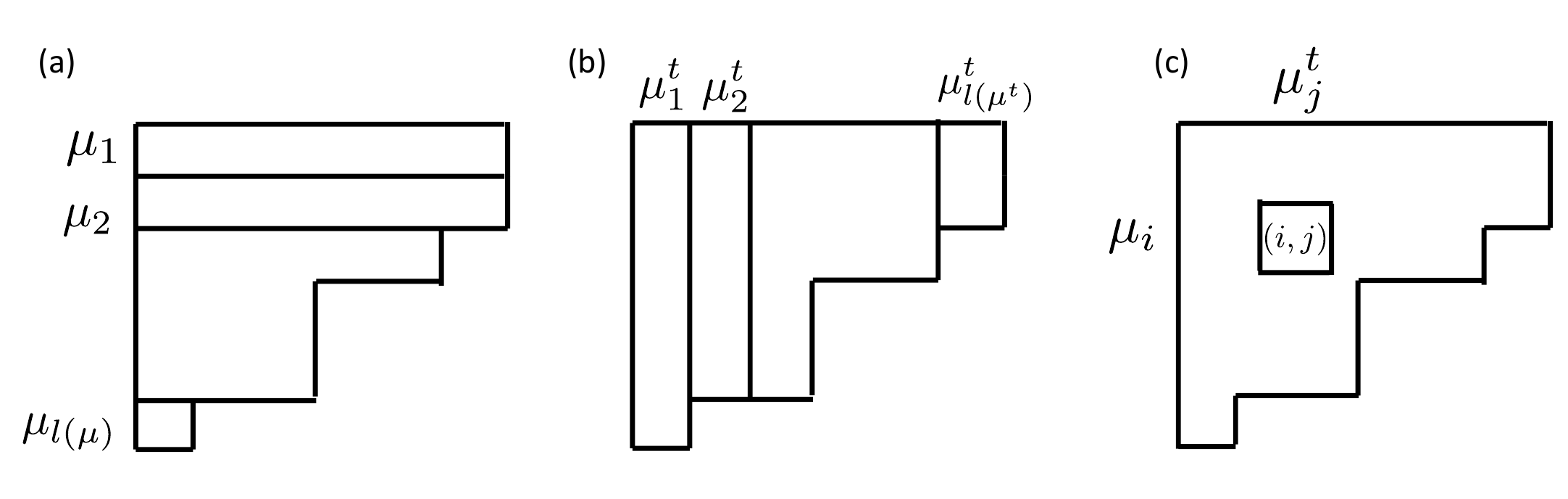}
    \caption{The Young diagram.  We define $\mu_{i}$ as (a), $\mu^{t}_{j}$ as (b), and the coordinates $(i,j)$ as (c). $\mu_{i}$ is the number of boxes in the $i$-th horizontal line. $\mu_{j}^{t}$ is the number of boxes in the $j$-th vertical line.}
    \label{Young}
\end{figure}

\paragraph{Schur function}
\par~\par
\noindent
We define the Schur function $s_{\mu}(x_{1},x_{2},...,x_{N})$ and the skew Schur function $s_{\mu/\nu}(x_{1},x_{2},...,x_{N})$ as follows:
\begin{align} 
s_{\mu}(x_{1},x_{2},...,x_{N}) &:= \frac{\mathrm{det} \lp x_{j}^{l_{i}+N-i} \rp}{\mathrm{det} \lp x_{j}^{N-i} \rp}, \\
s_{\mu/\nu}(x_{1},x_{2},...,x_{N}) &:= \sum_{\lambda}N^{\mu}_{\nu \lambda}s_{\lambda},
\end{align}
where $N^{\mu}_{\nu \lambda}$ is the Littlewood-Richardson coefficient. The skew schur function has the following properties:
\begin{eqnarray}
s_{\lambda/\mu}(\alpha \bold{x}) &=& \alpha^{|\lambda|-|\mu|}s_{\lambda/\mu}(\bold{x})
\label{schur1}
 \\
\sum_{\eta}s_{\eta/\lambda}(\bold{x})s_{\eta/\mu}(\bold{y}) &=& \prod_{i,j=1}^{\infty}(1-x_{i}y_{j})^{-1}\sum_{\tau}s_{\mu/\tau}(\bold{x})s_{\lambda/\tau}(\bold{y})
\label{schur2}
 \\
\sum_{\eta}s_{\eta^{t}/\lambda}(\bold{x})s_{\eta/\mu}(\bold{y}) &=& \prod_{i,j=1}^{\infty}(1+x_{i}y_{j})\sum_{\tau}s_{\mu^{t}/\tau}(\bold{x})s_{\lambda^{t}/\tau^{t}}(\bold{y})
\label{schur3}
\end{eqnarray}

\paragraph{Some formulae}
\par~\par
\noindent
We also provide some useful formulae to calculate the partition function:
\begin{eqnarray}
\prod_{i,j=1}^{\infty} \frac{1-Qq^{\nu_{i}-j}t^{\mu_{j}^{t}-i+1}}{1-Qq^{-j}t^{-i+1}} &=& \prod_{(i,j) \in \nu}(1-Qq^{\nu_{i}-j}t^{\mu_{j}^{t}-i+1})\prod_{(i,j) \in \mu}(1-Qq^{-\mu_{i}+j-1}t^{-\nu_{j}^{t}+i}) ~~~~~~~~~~~~\\
\prod_{i,j=1}^{\infty} \frac{1-Qt^{\nu_{j}^{t}-i+\frac{1}{2}}q^{-j+\frac{1}{2}}}{1-Qt^{-i+\frac{1}{2}}q^{-j+\frac{1}{2}}} &=& \prod_{(i,j) \in \nu}(1-Qq^{-j+\frac{1}{2}}t^{i-\frac{1}{2}}) \\
\prod_{i,j=1}^{\infty} \frac{1-Qq^{\nu_{i}-j+\frac{1}{2}}t^{-i+\frac{1}{2}}}{1-Qq^{-j+\frac{1}{2}}t^{-i+\frac{1}{2}}} &=&\prod_{(i,j) \in \nu}(1-Qq^{j-\frac{1}{2}}t^{-i+\frac{1}{2}})
\end{eqnarray}

\paragraph{Theta function}
\par~\par
\noindent
We define the theta function as follows:
\begin{eqnarray}
\theta_{1}(\tau ; x) :=
-\mathrm{i} e^{\frac{\mathrm{i} \pi \tau}{4}}x^{\frac{1}{2}}
\prod_{n=0}^{\infty}
\Bigl\{
(1-e^{2 \pi \mathrm{i}(n+1) \tau})
(1-e^{2 \pi \mathrm{i}(n+1) \tau}x)
(1-e^{2 \pi \mathrm{i} n \tau}x^{-1})
\Bigr\}
.
\end{eqnarray}
We sometimes write the theta function $\theta_{1}(\tau ; z)$ as follows:
\begin{eqnarray}
\theta_1 ( \tau | z ) :=
 -\mathrm{i} e^{\frac{\mathrm{i} \pi \tau}{4}}e^{\mathrm{i} \pi z}
\prod_{n=0}^{\infty} 
\Bigl\{
(1-e^{2 \pi \mathrm{i}(n+1) \tau})
(1-e^{2 \pi \mathrm{i}(n+1) \tau}e^{2 \pi \mathrm{i} z})
(1-e^{2 \pi i n \tau}e^{-2 \pi \mathrm{i} z})
 \Bigr\}.
\end{eqnarray}
This theta function has the following property:
\begin{eqnarray}
 \theta_{1}(\tau | - z ) = -\theta_{1}(\tau | z ).
\end{eqnarray}

\subsection{The refined topological vertex}
\begin{figure}[t]
\centering
    \includegraphics[width=5cm]{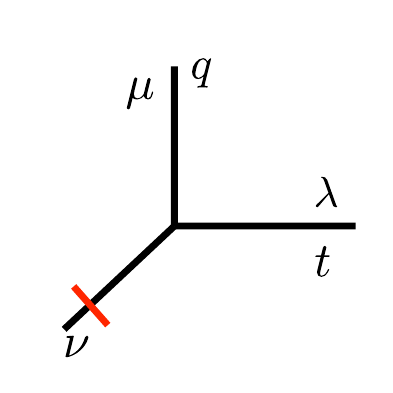}
    \caption{The pictorial description of the refined topological vertex.}
    \label{vertex}
\end{figure}

We define the refined topological vertex $C_{\lambda \mu \nu}(t,q)$ as
\begin{align}
C_{\lambda \mu \nu}(t,q) &:= t^{-\frac{||\mu^{t}||^{2}}{2}}q^{\frac{||\mu||^2 + ||\nu||^{2}}{2}} \tilde{Z}_{\nu}(t,q)
 \sum_{\eta}\Bigl(\frac{q}{t}\Bigr)^{\frac{|\eta| + |\lambda| - |\mu|}{2}}  s_{\lambda^{t}/\eta}(t^{-\rho}q^{-\nu})s_{\mu/\eta}(t^{-\nu^{t}}q^{-\rho}), \\
\tilde{Z}_{\nu}(t,q) &:= \prod_{(i,j) \in \nu}(1-q^{\nu_{i}-j}t^{\nu_{j}^{t} -i +1})^{-1},
\end{align}
where $|\mu|$ is the total number of boxes in the Young diagram $\mu$. $||\mu||$ and $\rho$ are defined as follows:
\begin{eqnarray}
||\mu||:=\sum_{i=1}^{l(\mu)}\mu^{2}_{i}, \hspace{2em}
\rho:=-i+\frac{1}{2}.
\end{eqnarray}
We can calculate the partition function of the refined topological string on a non-compact toric Calabi-Yau manifold. This can be done by choosing the preferred direction (a red line in Fig.\ref{vertex}) in the web diagram. Note that the partition function should be independent of the preferred direction.

\subsection{M-strigns calculation}
In this subsection, we provide the calculation details of the partition function of the M-strings. We first calculate the building block defined in Section \ref{Mstrings}.
\begin{align}
\mathcal{Z}^{\text{build}}_{\mu_{1}\mu_{2}\nu_{1}\nu_{2}}
&=
\sum_{\eta_{1,2},\tilde{\eta}_{1,2}}
(-Q_{1})^{|\eta_{1}|}(-Q_{2})^{|\eta_{2}|}(-\tilde{Q}_{1})^{|\tilde{\eta}_{1}|}(-\tilde{Q}_{2})^{|\tilde{\eta}_{2}|}
C_{\tilde{\eta}_{2}\eta_{1}^{t}\mu_{1}}(t,q)C_{\tilde{\eta}_{1}^{t}\eta_{1}\nu_{1}^{t}}(q,t)
C_{\tilde{\eta}_{1}\eta_{2}^{t}\mu_{2}}(t,q)C_{\tilde{\eta}_{2}^{t}\eta_{2}\nu_{2}^{t}}(q,t)
\nonumber \\
&=
q^{\frac{ ||\mu_{1}||^{2} + ||\mu_{2}||^{2}}{2}}t^{\frac{ ||\nu_{1}^{t}||^{2} + ||\nu_{2}^{t}||^{2}}{2}}
\tilde{Z}_{\mu_{1}}(t,q)\tilde{Z}_{\nu_{1}^{t}}(q,t) \tilde{Z}_{\mu_{2}}(t,q)\tilde{Z}_{\nu_{2}^{t}}(q,t)
\nonumber \\
&\hspace{1em} \times
\sum_{
{\scriptsize
\begin{matrix}
\eta_{1,2},\tilde{\eta}_{1,2} \\
\xi_{1,2},\tilde{\xi}_{1,2}
\end{matrix}}
}
(-Q_{1})^{|\eta_{1}|}(-Q_{2})^{|\eta_{2}|}(-\tilde{Q}_{1})^{|\tilde{\eta}_{1}|}(-\tilde{Q}_{2})^{|\tilde{\eta}_{2}|}
\Bigl(\frac{q}{t}\Bigr)^{\frac{|\xi_{1}| -|\tilde{\xi}_{1}| + |\xi_{2}| -|\tilde{\xi}_{2}|}{2}}
\nonumber \\
&\hspace{1em} \times
s_{\tilde{\eta}^{t}_{2}/\xi_{1}}(t^{-\rho}q^{-\mu_{1}})
s_{\eta_{1}^{t}/\xi_{1}}(t^{-\mu_{1}^{t}}q^{-\rho})
s_{\tilde{\eta}_{1}/\tilde{\xi}_{1}}(q^{-\rho}t^{-\nu_{1}^{t}})
s_{\eta_{1}/\tilde{\xi}_{1}}(q^{-\nu_{1}}t^{-\rho})
\nonumber \\
&\hspace{1em} \times
s_{\tilde{\eta}_{1}^{t}/\xi_{2}}(t^{-\rho}q^{-\mu_{2}})
s_{\eta_{2}^{t}/\xi_{2}}(t^{-\mu_{2}^{t}}q^{-\rho})
s_{\tilde{\eta}_{2}/\tilde{\xi}_{2}}(q^{-\rho}t^{-\nu_{2}^{t}})
s_{\eta_{2}/\tilde{\xi}_{2}}(q^{-\nu_{2}}t^{-\rho}).
\end{align}
Thus, we have to calculate
\begin{eqnarray}
G(\alpha_{1,2},\tilde{\alpha}_{1,2},\beta_{1,2},\tilde{\beta}_{1,2};\bold{x})
&:=&
\sum_{
{\scriptsize
\begin{matrix}
\eta_{1,2},\tilde{\eta}_{1,2} \\
\xi_{1,2},\tilde{\xi}_{1,2}
\end{matrix}}
}
\alpha_{1}^{|\eta_{1}|}\alpha_{2}^{|\eta_{2}|}
\tilde{\alpha}_{1}^{|\tilde{\eta}_{1}|}\tilde{\alpha}_{2}^{|\tilde{\eta}_{2}|}
\beta_{1}^{|\xi_{1}|}\tilde{\beta}_{1}^{|\tilde{\xi}_{1}|}
\beta_{2}^{|\xi_{2}|}\tilde{\beta}_{2}^{|\tilde{\xi}_{2}|}
\nonumber \\ &&\times
s_{\tilde{\eta}^{t}_{2}/\xi_{1}}(x_{1})
s_{\eta_{1}^{t}/\xi_{1}}(x_{2})
s_{\tilde{\eta}_{1}/\tilde{\xi}_{1}}(x_{3})
s_{\eta_{1}/\tilde{\xi}_{1}}(x_{4})
\nonumber \\ &&\times
s_{\tilde{\eta}_{1}^{t}/\xi_{2}}(x_{5})
s_{\eta_{2}^{t}/\xi_{2}}(x_{6})
s_{\tilde{\eta}_{2}/\tilde{\xi}_{2}}(x_{7})
s_{\eta_{2}/\tilde{\xi}_{2}}(x_{8}),
\end{eqnarray}
where we define
\begin{eqnarray}
&& \alpha_{1,2}:=-Q_{1,2},~\tilde{\alpha}_{1,2}:=-\tilde{Q}_{1,2},
\nonumber \\
&& \beta_{1,2}:=\sqrt{\frac{q}{t}},~\tilde{\beta}_{1,2}:=\sqrt{\frac{t}{q}},
\nonumber \\
&& x^{i}_{1}:=t^{i-\frac{1}{2}}q^{-\mu_{1,i}},~x^{i}_{2}:=t^{-\mu_{1,i}^{t}}q^{i-\frac{1}{2}},~~
 x^{i}_{3}:=t^{-\nu_{1,i}^{t}}q^{i-\frac{1}{2}},~x^{i}_{4}:=t^{i-\frac{1}{2}}q^{-\nu_{1,i}},~~
\nonumber \\
&& x^{i}_{5}:=t^{i-\frac{1}{2}}q^{-\mu_{2,i}},~x^{i}_{6}:=t^{-\mu_{2,i}^{t}}q^{i-\frac{1}{2}},~~
 x^{i}_{7}:=t^{-\nu_{2,i}^{t}}q^{i-\frac{1}{2}},~x^{i}_{8}:=t^{i-\frac{1}{2}}q^{-\nu_{2,i}}.
\label{parameter}
\end{eqnarray}
Then, by using the formulae (\ref{schur1}), (\ref{schur2}), and (\ref{schur3}), we obtain the following recursion formula:
\begin{eqnarray}
&&G(\alpha_{1,2},\tilde{\alpha}_{1,2},\beta_{1,2},\tilde{\beta}_{1,2};\bold{x})
\nonumber \\
&&=
\prod_{i,j=1}^{\infty}
\frac{
(1+\tilde{\alpha}_{2}x_{1}^{i}x_{7}^{j})
(1+\alpha_{1}x_{2}^{i}x_{4}^{j})
(1+\tilde{\alpha}_{1}x_{3}^{i}x_{5}^{j})
(1+\alpha_{2}x_{6}^{i}x_{8}^{j})
}{
(1-\tilde{\alpha}_{1}^{-1}\alpha_{1}^{-1}\tilde{\beta}_{2}Ax_{1}^{i}x_{6}^{j})
(1-\tilde{\alpha}_{1}\alpha_{1}\tilde{\beta}_{1}x_{2}^{i}x_{5}^{j})
(1-\tilde{\alpha}_{2}^{-1}\alpha_{1}^{-1}\beta_{2}Ax_{3}^{i}x_{8}^{j})
(1-\tilde{\alpha}_{2}\alpha_{1}\beta_{1}x_{4}^{i}x_{7}^{j})
}
\nonumber \\ &&\times
\frac{
(1+\alpha_{1}^{-1}\beta_{2}\tilde{\beta}_{2}Ax_{1}^{i}x_{3}^{j})
(1+\tilde{\alpha}_{2}^{-1}\beta_{2}\tilde{\beta}_{1}Ax_{2}^{i}x_{8}^{j})
(1+\tilde{\alpha}_{1}^{-1}\beta_{1}\tilde{\beta}_{2}Ax_{4}^{i}x_{6}^{j})
(1+\alpha_{2}^{-1}\beta_{1}\tilde{\beta}_{1}Ax_{5}^{i}x_{7}^{j})
}{
(1-AB\beta_{1}^{-1}x^{i}_{1}x^{j}_{2})
(1-AB\tilde{\beta}_{1}^{-1}x^{i}_{3}x^{j}_{4})
(1-AB\beta_{2}^{-1}x^{i}_{5}x^{j}_{6})
(1-AB\tilde{\beta}_{2}^{-1}x^{i}_{7}x^{j}_{8})
}
\nonumber \\ &&\times
\frac{
(1+\tilde{\alpha}_{2}ABx^{i}_{1}x^{j}_{7})
(1+\alpha_{1}ABx^{i}_{2}x^{j}_{4})
(1+\tilde{\alpha}_{1}ABx^{i}_{3}x^{j}_{5})
(1+\alpha_{2}ABx^{i}_{6}x^{j}_{8})
}{
(1-\tilde{\alpha}_{1}^{-1}\alpha_{1}^{-1}\tilde{\beta}_{2}A^{2}Bx_{1}^{i}x_{6}^{j})
(1-\tilde{\alpha}_{1}\alpha_{1}\tilde{\beta}_{1}ABx_{2}^{i}x_{5}^{j})
(1-\tilde{\alpha}_{2}^{-1}\alpha_{1}^{-1}\beta_{2}A^{2}Bx_{3}^{i}x_{8}^{j})
(1-\tilde{\alpha}_{2}\alpha_{1}\beta_{1}ABx_{4}^{i}x_{7}^{j})
}
\nonumber \\ &&\times
\frac{
(1+\alpha_{2}^{-1}\beta_{1}\tilde{\beta}_{1}A^{2}Bx_{5}^{i}x_{7}^{j})
(1+\alpha_{1}^{-1}\beta_{2}\tilde{\beta}_{2}A^{2}Bx_{1}^{i}x_{3}^{j})
(1+\tilde{\alpha}_{2}^{-1}\beta_{2}\tilde{\beta}_{1}A^{2}Bx_{2}^{i}x_{8}^{j})
(1+\tilde{\alpha}_{1}^{-1}\beta_{1}\tilde{\beta}_{2}A^{2}Bx_{4}^{i}x_{6}^{j})
}{
(1-A^{2}B^{2}\beta_{1}^{-1}x^{i}_{1}x^{j}_{2})
(1-A^{2}B^{2}\tilde{\beta}_{1}^{-1}x^{i}_{3}x^{j}_{4})
(1-A^{2}B^{2}\beta_{2}^{-1}x^{i}_{5}x^{j}_{6})
(1-A^{2}B^{2}\tilde{\beta}_{2}^{-1}x^{i}_{7}x^{j}_{8})
}
\nonumber \\ &&\times
G(\alpha_{1,2},\tilde{\alpha}_{1,2},\beta_{1,2},\tilde{\beta}_{1,2};AB\bold{x}),
\end{eqnarray}
where we define
\begin{eqnarray}
A:=\alpha_{1}\alpha_{2}\tilde{\alpha}_{1}\tilde{\alpha}_{2},~
B:=\beta_{1}\beta_{2}\tilde{\beta}_{1}\tilde{\beta}_{2}.
\end{eqnarray}
Then, we obtain
\begin{align}
G(\alpha_{1,2},\tilde{\alpha}_{1,2},\beta_{1,2},\tilde{\beta}_{1,2};\bold{x})
&=
\prod_{n=0}^{2M-1}\prod_{i,j=1}^{\infty}
\Biggl\{
\frac{
(1+\tilde{\alpha}_{2}A^{n}B^{n}x_{1}^{i}x_{7}^{j})
(1+\alpha_{1}A^{n}B^{n}x_{2}^{i}x_{4}^{j})
}{
(1-\tilde{\alpha}_{1}^{-1}\alpha_{1}^{-1}\tilde{\beta}_{2}A^{n+1}B^{n}x_{1}^{i}x_{6}^{j})
(1-\tilde{\alpha}_{1}\alpha_{1}\tilde{\beta}_{1}A^{n}B^{n}x_{2}^{i}x_{5}^{j})
}
\nonumber \\
&\hspace{1em} \times
\frac{
(1+\tilde{\alpha}_{1}A^{n}B^{n}x_{3}^{i}x_{5}^{j})
(1+\alpha_{2}A^{n}B^{n}x_{6}^{i}x_{8}^{j})
}{
(1-\tilde{\alpha}_{2}^{-1}\alpha_{1}^{-1}\beta_{2}A^{n+1}B^{n}x_{3}^{i}x_{8}^{j})
(1-\tilde{\alpha}_{2}\alpha_{1}\beta_{1}A^{n}B^{n}x_{4}^{i}x_{7}^{j})
}
\nonumber \\
&\hspace{1em} \times
\frac{
(1+\alpha_{1}^{-1}\beta_{2}\tilde{\beta}_{2}A^{n+1}B^{n}x_{1}^{i}x_{3}^{j})
(1+\tilde{\alpha}_{2}^{-1}\beta_{2}\tilde{\beta}_{1}A^{n+1}B^{n}x_{2}^{i}x_{8}^{j})
}{
(1-A^{n+1}B^{n+1}\beta_{1}^{-1}x^{i}_{1}x^{j}_{2})
(1-A^{n+1}B^{n+1}\tilde{\beta}_{1}^{-1}x^{i}_{3}x^{j}_{4})
}
\nonumber \\
&\hspace{1em} \times
\frac{
(1+\tilde{\alpha}_{1}^{-1}\beta_{1}\tilde{\beta}_{2}A^{n+1}B^{n}x_{4}^{i}x_{6}^{j})
(1+\alpha_{2}^{-1}\beta_{1}\tilde{\beta}_{1}A^{n+1}B^{n}x_{5}^{i}x_{7}^{j})
}{
(1-A^{n+1}B^{n+1}\beta_{2}^{-1}x^{i}_{5}x^{j}_{6})
(1-A^{n+1}B^{n+1}\tilde{\beta}_{2}^{-1}x^{i}_{7}x^{j}_{8})
}
\Biggr\} \notag \\
&\hspace{1em} \times
G(\alpha_{1,2},\tilde{\alpha}_{1,2},\beta_{1,2},\tilde{\beta}_{1,2};A^M B^M \bold{x}).
\end{align}
\par
Next, we consider the limit $M\to\infty$. The non-trivial contributions in $G$ appear if and only if the following conditions are satisfied under the assumption $\lim_{M\to\infty}A^{M}=0$:
\begin{eqnarray}
\eta_{1}=\eta_{2}=\tilde{\eta}_{1}=\tilde{\eta}_{2}
=
\xi_{1}=\xi_{2}=\tilde{\xi}_{1}=\tilde{\xi}_{2}.
\end{eqnarray}
From the definition of the skew Schur function with this condition, the limit $M\to\infty$ simplifies $G$ as
\begin{align} %
\lim_{M \to \infty}
G(\alpha_{1,2},\tilde{\alpha}_{1,2},\beta_{1,2},\tilde{\beta}_{1,2};A^M B^M \bold{x})
=
\sum_{\mu} A^{|\mu|} B^{|\mu|}
=
\prod_{k = 1}^{\infty}
{1 \over 1 - A^k B^k}.
\end{align}
Thus, we obtain
\begin{align}
G(\alpha_{1,2},\tilde{\alpha}_{1,2},\beta_{1,2},\tilde{\beta}_{1,2};\bold{x})
&=
\prod_{k=1}^{\infty}(1-A^{k}B^{k})^{-1}
\nonumber \\
&\hspace{1em} \times
\prod_{n=0}^{2M-1}\prod_{i,j=1}^{\infty}
\Biggl\{
\frac{
(1+\alpha_{1}A^{n}B^{n}x_{2}^{i}x_{4}^{j})
(1+\alpha_{2}A^{n}B^{n}x_{6}^{i}x_{8}^{j})
}{
(1-\tilde{\alpha}_{2}\alpha_{1}\beta_{1}A^{n}B^{n}x_{4}^{i}x_{7}^{j})
(1-\tilde{\alpha}_{2}^{-1}\alpha_{1}^{-1}\beta_{2}A^{n+1}B^{n}x_{3}^{i}x_{8}^{j})
}
\nonumber \\
&\hspace{1em} \times
\frac{
(1+\tilde{\alpha}_{1}A^{n}B^{n}x_{3}^{i}x_{5}^{j})
(1+\tilde{\alpha}_{2}A^{n}B^{n}x_{1}^{i}x_{7}^{j})
}{
(1-\tilde{\alpha}_{1}\alpha_{1}\tilde{\beta}_{1}A^{n}B^{n}x_{2}^{i}x_{5}^{j})
(1-\tilde{\alpha}_{1}^{-1}\alpha_{1}^{-1}\tilde{\beta}_{2}A^{n+1}B^{n}x_{1}^{i}x_{6}^{j})
}
\nonumber \\
&\hspace{1em} \times
\frac{
(1+\alpha_{2}^{-1}\beta_{1}\tilde{\beta}_{1}A^{n+1}B^{n}x_{5}^{i}x_{7}^{j})
(1+\alpha_{1}^{-1}\beta_{2}\tilde{\beta}_{2}A^{n+1}B^{n}x_{1}^{i}x_{3}^{j})
}{
(1-A^{n+1}B^{n+1}\beta_{2}^{-1}x_{5}x_{6})
(1-A^{n+1}B^{n+1}\beta_{1}^{-1}x_{1}x_{2})
}
\nonumber \\
&\hspace{1em} \times
\frac{
(1+\tilde{\alpha}_{2}^{-1}\beta_{2}\tilde{\beta}_{1}A^{n+1}B^{n}x_{2}^{i}x_{8}^{j})
(1+\tilde{\alpha}_{1}^{-1}\beta_{1}\tilde{\beta}_{2}A^{n+1}B^{n}x_{4}^{i}x_{6}^{j})
}{
(1-A^{n+1}B^{n+1}\tilde{\beta}_{2}^{-1}x_{7}x_{8})
(1-A^{n+1}B^{n+1}\tilde{\beta}_{1}^{-1}x_{3}x_{4})
}
\Biggr\}.
\end{align}
Therefore, by using the relations (\ref{parameter}), we obtain
\begin{eqnarray}
\mathcal{Z}^{\text{build}}_{\mu_{1}\mu_{2}\nu_{1}\nu_{2}}
&=&
\prod_{k=1}^{\infty}(1-Q_{\tau})^{-1}
\nonumber \\ &&\times
\prod_{n=0}^{\infty}\prod_{i,j=1}^{\infty}
\frac{
(1-Q_{1}Q_{\tau}^{n}t^{-\mu_{1,j}^{t}+i-\frac{1}{2}}q^{-\nu_{1,i}+j-\frac{1}{2}})
(1-Q_{2}Q_{\tau}^{n}t^{-\mu_{2,j}^{t}+i-\frac{1}{2}}q^{-\nu_{2,i}+j-\frac{1}{2}})
}{
(1-\tilde{Q}_{2}Q_{1}Q_{\tau}^{n}t^{-\nu_{2,j}^{t}+i-1}q^{-\nu_{1,i}+j})
(1-\tilde{Q}_{2}^{-1}Q_{1}^{-1}Q_{\tau}^{n+1}t^{-\nu_{1,j}^{t}+i-1}q^{-\nu_{2,i}+j})
}
\nonumber \\ &&\times
\frac{
(1-\tilde{Q}_{1}Q_{\tau}^{n}t^{-\nu_{1,j}^{t}+i-\frac{1}{2}}q^{-\mu_{2,i}+j-\frac{1}{2}})
(1-\tilde{Q}_{2}Q_{\tau}^{n}t^{-\nu_{2,j}^{t}+i-\frac{1}{2}}q^{-\mu_{1,i}+j-\frac{1}{2}})
}{
(1-\tilde{Q}_{1}Q_{1}Q_{\tau}^{n}t^{-\mu_{1,j}^{t}+i}q^{-\mu_{2,i}+j-1})
(1-\tilde{Q}_{1}^{-1}Q_{1}^{-1}Q_{\tau}^{n+1}t^{-\mu_{2,j}^{t}+i}q^{-\mu_{1,i}+j-1})
}
\nonumber \\ &&\times
\frac{
(1-Q_{2}^{-1}Q_{\tau}^{n+1}t^{-\nu_{2,j}^{t}+i-\frac{1}{2}}q^{-\mu_{2,i}+j-\frac{1}{2}})
(1-Q_{1}^{-1}Q_{\tau}^{n+1}t^{-\nu_{1,j}^{t}+i-\frac{1}{2}}q^{-\mu_{1,i}+j-\frac{1}{2}})
}{
(1-Q_{\tau}^{n+1}t^{-\mu_{2,j}^{t}+i}q^{-\mu_{2,i}+j-1})
(1-Q_{\tau}^{n+1}t^{-\mu_{1,j}^{t}+i}q^{-\mu_{1,i}+j-1})
}
\nonumber \\ &&\times
\frac{
(1-\tilde{Q}_{2}^{-1}Q_{\tau}^{n+1}t^{-\mu_{1,j}^{t}+i-\frac{1}{2}}q^{-\nu_{2,i}+j-\frac{1}{2}})
(1-\tilde{Q}_{1}^{-1}Q_{\tau}^{n+1}t^{-\mu_{2,j}^{t}+i-\frac{1}{2}}q^{-\nu_{1,i}+j-\frac{1}{2}})
}{
(1-Q_{\tau}^{n+1}t^{-\nu_{2,j}^{t}+i-1}q^{-\nu_{2,i}+j})
(1-Q_{\tau}^{n+1}t^{-\nu_{1,j}^{t}+i-1}q^{-\nu_{1,i}+j})
},
\nonumber \\
\end{eqnarray}
where we define $Q_{\tau}:=Q_{1}Q_{2}\tilde{Q}_{1}\tilde{Q}_{2}$.
\par
Finally, by dividing $\mathcal{Z}_{\emptyset \emptyset\nu_{1}\nu_{2}}^{\text{build}}$ and $\mathcal{Z}_{\mu_{1}\mu_{2}\emptyset \emptyset}^{\text{build}}$ by $\mathcal{Z}_{\emptyset \emptyset\emptyset \emptyset}^{\text{build}}$, we obtain
\begin{eqnarray}
\hat{\mathcal{Z}}_{\emptyset\emptyset\nu_{1}\nu_{2}}^{\text{build}}
&:=&
\mathcal{Z}^{\text{build}}_{\emptyset\emptyset\nu_{1}\nu_{2}}/\mathcal{Z}_{\emptyset \emptyset\emptyset \emptyset}^{\text{build}}
\nonumber \\
&=&
\prod_{n=0}^{\infty}
\prod_{(i,j) \in \nu_{1}}
\frac{
(1-Q_{1}Q_{\tau}^{n}t^{j-\frac{1}{2}}q^{-\nu_{1,i}+j-\frac{1}{2}})
(1-Q_{1}^{-1}Q_{\tau}^{n+1}t^{-j+\frac{1}{2}}q^{\nu_{1,i}-j+\frac{1}{2}})
}{
(1-\tilde{Q}_{2}Q_{1}Q_{\tau}^{n}t^{-\nu_{2,j}^{t}+i-1}q^{-\nu_{1,i}+j})
(1-\tilde{Q}_{2}^{-1}Q_{1}^{-1}Q_{\tau}^{n+1}t^{\nu_{2,j}^{t}-i}q^{\nu_{1,i}-j+1})
}
\nonumber \\ &&\times
\frac{
(1-\tilde{Q}_{1}Q_{\tau}^{n}t^{-j+\frac{1}{2}}q^{\nu_{1,i}-j+\frac{1}{2}})
(1-\tilde{Q}_{1}^{-1}Q_{\tau}^{n+1}t^{j-\frac{1}{2}}q^{-\nu_{1,i}+j-\frac{1}{2}})
}{
(1-Q_{\tau}^{n+1}t^{-\nu_{1,j}^{t}+i-1}q^{-\nu_{1,i}+j})
(1-Q_{\tau}^{n+1}t^{\nu_{1,j}^{t}-i}q^{\nu_{1,i}-j+1})
}
\nonumber \\ &&\times
\prod_{(i,j) \in \nu_{2}}
\frac{
(1-Q_{2}Q_{\tau}^{n}t^{j-\frac{1}{2}}q^{-\nu_{2,i}+j-\frac{1}{2}})
(1-Q_{2}^{-1}Q_{\tau}^{n+1}t^{-j+\frac{1}{2}}q^{\nu_{2,i}-j+\frac{1}{2}})
}{
(1-\tilde{Q}_{2}Q_{1}Q_{\tau}^{n}t^{\nu_{1,j}^{t}-i}q^{\nu_{2,i}-j+1})
(1-\tilde{Q}_{2}^{-1}Q_{1}^{-1}Q_{\tau}^{n+1}t^{-\nu_{1,j}^{t}+i-1}q^{-\nu_{2,i}+j})
}
\nonumber \\ &&\times
\frac{
(1-\tilde{Q}_{2}Q_{\tau}^{n}t^{-j+\frac{1}{2}}q^{\nu_{2,i}-j+\frac{1}{2}})
(1-\tilde{Q}_{2}^{-1}Q_{\tau}^{n+1}t^{j-\frac{1}{2}}q^{-\nu_{2,i}+j-\frac{1}{2}})
}{
(1-Q_{\tau}^{n+1}t^{-\nu_{2,j}^{t}+i-1}q^{-\nu_{2,i}+j})
(1-Q_{\tau}^{n+1}t^{\nu_{2,j}^{t}-i}q^{\nu_{2,i}-j+1})
},
\nonumber \\
\hat{\mathcal{Z}}_{\mu_{1}\mu_{2}\emptyset\emptyset}^{\text{build}}
&:=&
\mathcal{Z}_{\mu_{1}\mu_{2}\emptyset \emptyset}^{\text{build}}/\mathcal{Z}_{\emptyset \emptyset\emptyset \emptyset}^{\text{build}}
\nonumber \\
&=&
\prod_{n=0}^{\infty}
\prod_{(i,j) \in \mu_{1}}
\frac{
(1-Q_{1}Q_{\tau}^{n}t^{-j+\frac{1}{2}}q^{\mu_{1,i}-j+\frac{1}{2}})
(1-Q_{1}^{-1}Q_{\tau}^{n+1}t^{j-\frac{1}{2}}q^{-\mu_{1,i}+j-\frac{1}{2}})
}{
(1-\tilde{Q}_{1}Q_{1}Q_{\tau}^{n}t^{\mu_{2,j}^{t}-i+1}q^{\mu_{1,i}-j})
(1-\tilde{Q}_{1}^{-1}Q_{1}^{-1}Q_{\tau}^{n+1}t^{-\mu_{2,j}^{t}+i}q^{-\mu_{1,i}+j-1})
}
\nonumber \\ &&\times
\frac{
(1-\tilde{Q}_{2}Q_{\tau}^{n}t^{j-\frac{1}{2}}q^{-\mu_{1,i}+j-\frac{1}{2}})
(1-\tilde{Q}_{2}^{-1}Q_{\tau}^{n+1}t^{-j+\frac{1}{2}}q^{\mu_{1,i}-j+\frac{1}{2}})
}{
(1-Q_{\tau}^{n+1}t^{-\mu_{1,j}^{t}+i}q^{-\mu_{1,i}+j-1})
(1-Q_{\tau}^{n+1}t^{\mu_{1,j}^{t}-i+1}q^{\mu_{1,i}-j})
}
\nonumber \\ &&\times
\prod_{(i,j) \in \mu_{2}}
\frac{
(1-Q_{2}Q_{\tau}^{n}t^{-j+\frac{1}{2}}q^{\mu_{2,i}-j+\frac{1}{2}})
(1-Q_{2}^{-1}Q_{\tau}^{n+1}t^{j-\frac{1}{2}}q^{-\mu_{2,i}+j-\frac{1}{2}})
}{
(1-\tilde{Q}_{1}Q_{1}Q_{\tau}^{n}t^{-\mu_{1,j}^{t}+i}q^{-\mu_{2,i}+j-1})
(1-\tilde{Q}_{1}^{-1}Q_{1}^{-1}Q_{\tau}^{n+1}t^{\mu_{1,i}^{t}-j+1}q^{\mu_{2,i}-j})
}
\nonumber \\ &&\times
\frac{
(1-\tilde{Q}_{1}Q_{\tau}^{n}t^{i-\frac{1}{2}}q^{-\mu_{2,i}+j-\frac{1}{2}})
(1-\tilde{Q}_{1}^{-1}Q_{\tau}^{n+1}t^{-i+\frac{1}{2}}q^{\mu_{2,i}-j+\frac{1}{2}})
}{
(1-Q_{\tau}^{n+1}t^{-\mu_{2,j}^{t}+i}q^{-\mu_{2,i}+j-1})
(1-Q_{\tau}^{n+1}t^{\mu_{2,j}^{t}-i+1}q^{\mu_{2,i}-j})
}
\end{eqnarray}
By using the above results and the definition of the theta function, we can obtain the partition function of the M-strings.

\providecommand{\href}[2]{#2}\begingroup\raggedright\endgroup

\end{document}